\documentclass[preprint2]{pp7}

\input{pp7.h}

\usepackage{amsmath}
\usepackage{xspace}
\usepackage[utf8]{inputenc}
\usepackage[skins]{tcolorbox}
\usepackage[colorlinks]{hyperref}

\usepackage{savesym}
\savesymbol{tablenum}
\usepackage{siunitx}
\restoresymbol{SIX}{tablenum}

\usepackage{amsfonts}
\usepackage{natbib}
\usepackage[T1]{fontenc}
\usepackage{floatrow}
\usepackage{csquotes}
\usepackage{graphicx}
\usepackage{float}
\graphicspath{{./figures/}} 

\begin{document}

\definecolor{MyDarkBlue}{rgb}{0,0.08,0.5}
\definecolor{MyDarkRed}{rgb}{0.7,0.02,0.02}
\definecolor{MyDarkGreen}{rgb}{0.0,0.7,0.0}
\definecolor{britishracinggreen}{rgb}{0.0, 0.26, 0.15}
\definecolor{cadmiumgreen}{rgb}{0.0, 0.42, 0.24}


\newcommand{\kataoka}[1]{\textcolor{MyDarkRed}{[Akimasa: #1]}}
\newcommand{\til}[1]{\textcolor{MyDarkGreen}{[Til: #1]}}
\newcommand{\ik}[1]{\textcolor{blue}{[Inga: #1]}}
\newcommand{\ikt}[1]{\textcolor{blue}{#1}}
\newcommand{\anna}[1]{\textcolor{cadmiumgreen}{#1}}
\newcommand{\questions}[1]{\textcolor{cadmiumgreen}{#1}}
\newcommand{\update}[1]{\textbf{#1}}
\newcommand{\lic}[1]{\textcolor{magenta}{#1}}

\newtcolorbox{summarybox}{enhanced,width=24em,drop fuzzy shadow southwest,colframe=gray!8,colback=gray!8}

\hypersetup{
      linkcolor=MyDarkBlue,
      citecolor=MyDarkBlue,
      urlcolor=MyDarkBlue,
}

\sisetup{range-phrase={\ensuremath{-}}, range-units = single, mode=text}

\title{\textbf{\LARGE Setting the Stage for Planet Formation:\\Measurements and Implications of the Fundamental Disk Properties}}
\shorttitle{Global Disk Properties}

\author {\textbf{\large Anna Miotello}}
\affil{\small\it European Southern Observatory, Karl-Schwarzschild-Str. 2,85748 Garching, Germany - amiotell@eso.org}

\author {\textbf{\large Inga Kamp}}
\affil{\small\it Kapteyn Astronomical Institute, University of Groningen, Groningen, The Netherlands}

\author {\textbf{\large Tilman Birnstiel}}
\affil{\small\it University Observatory, Faculty of Physics, Ludwig-Maximilians-Universit\"at M\"unchen, Scheinerstr.~1, 81679 Munich, Germany}
\affil{\small\it Exzellenzcluster ORIGINS, Boltzmannstr. 2, D-85748 Garching, Germany}

\author {\textbf{\large L. Ilsedore Cleeves}}
\affil{\small\it Astronomy Department, University of Virginia, Charlottesville, VA 22904, USA}

\author {\textbf{\large Akimasa Kataoka}}
\affil{\small\it National Astronomical Observatory of Japan, Osawa 2-21-1, Mitaka, Tokyo 181-8588, Japan}


\newcommand{\amax}{\ensuremath{a_\mathrm{max}}\xspace}
\newcommand{\Sigd}{\ensuremath{\Sigma_\mathrm{d}}\xspace}
\newcommand{\Sigg}{\ensuremath{\Sigma_\mathrm{g}}\xspace}
\newcommand{\Ro}{\ensuremath{R_\mathrm{out}}\xspace}
\newcommand{\Rod}{\ensuremath{R_\mathrm{out,dust}}\xspace}
\newcommand{\Rog}{\ensuremath{R_\mathrm{out,gas}}\xspace}
\newcommand{\Td}{\ensuremath{T_\mathrm{dust}}\xspace}
\newcommand{\Tg}{\ensuremath{T_\mathrm{gas}}\xspace}
\newcommand{\Md}{\ensuremath{M_\mathrm{dust}}\xspace}
\newcommand{\Mg}{\ensuremath{M_\mathrm{gas}}\xspace}
\newcommand{\Vk}{\ensuremath{V_\mathrm{K}}\xspace}
\newcommand{\Ok}{\ensuremath{\Omega_\mathrm{K}}\xspace}
\newcommand{\kapabs}{\ensuremath{\kappa_\mathrm{abs}}\xspace}


\begin{abstract}
      \baselineskip = 11pt
      \leftskip = 1.5cm
      \rightskip = 1.5cm
      \parindent=1pc
      {\small \textbf{Abstract} -
      The field of planet formation is in an exciting era, where recent observations of disks around low- to intermediate-mass stars made with state of the art interferometers and high-contrast optical and IR facilities have revealed a diversity of substructures, some possibly planet-related.  It is therefore important to understand the physical and chemical nature of the protoplanetary building blocks, as well as their spatial distribution, to better understand planet formation. Since PPVI, the field has seen tremendous improvements in observational capabilities, enabling both surveys of large samples of disks and high resolution imaging studies of a few bright disks. Improvements in data quality and sample size have, however, opened up many fundamental questions about properties such as the mass budget of disks, its spatial distribution, and its radial extent. Moreover, the vertical structure of disks has been studied in greater detail with spatially resolved observations, providing new insights on vertical layering and temperature stratification, yet also bringing rise to questions about other properties, such as material transport and viscosity. Each one of these properties - disk mass, surface density distribution, outer radius, vertical extent, temperature structure, and transport - is of fundamental interest as they collectively set the stage for disk evolution and corresponding planet formation theories. In this chapter, we will review our understanding of the fundamental properties of disks including the relevant observational techniques to probe their nature, modeling methods, and the respective caveats. Finally, we discuss the implications for theories of disk evolution and planet formation underlining what new questions have since arisen as our observational facilities have improved.
      \\~\\~\\~}
\end{abstract}

\section{\textbf{INTRODUCTION}}

Protoplanetary disks around low- to intermediate-mass stars ($0.1 M_{\odot} \lesssim M_{\star} \lesssim 5 M_{\odot}$) are thought to harbor on-going planet formation and/or recently formed planets. This picture is now supported by observations of rich, potentially planet-related, substructure at both infrared and sub-mm wavelengths. Therefore, it is essential to understand the physical and chemical nature of the protoplanetary building-blocks and their spatial distribution to better understand planet formation mechanisms and the demographics of the resulting planetary systems.

\begin{figure*}[h!]
      \includegraphics[width=\hsize,trim={1.3cm 3.2cm .4cm 5.2cm},clip]{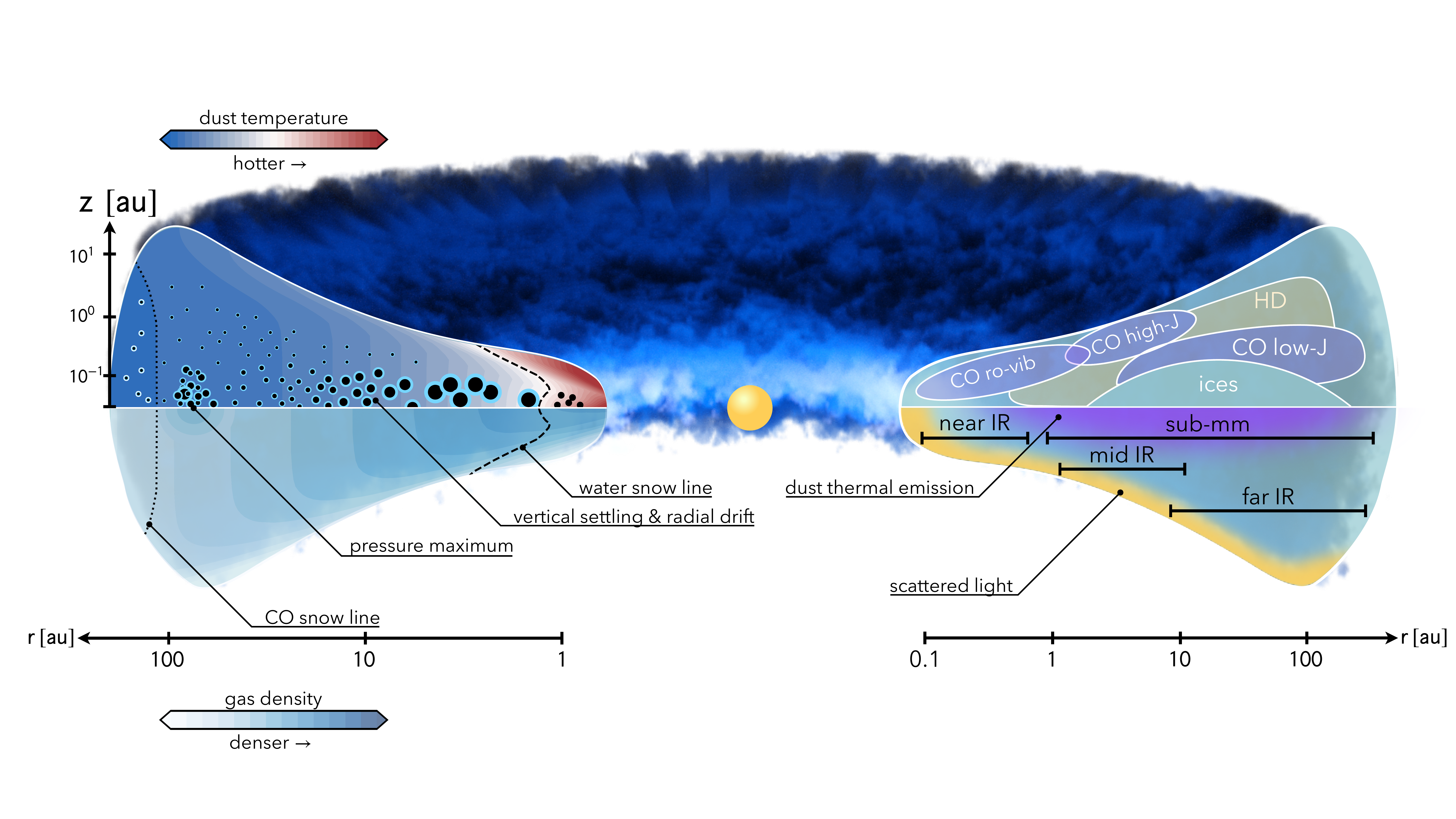}
      \caption{ \emph{Left} - Illustration of the dust temperature and gas density structure in the protoplanetary disks. A sketched distribution of dust particles is shown by the black circles, whose size variation is represented by the symbol size. Bare grains are present within the water snow line (dashed curve), H$_2$O-coated grains are contoured in blue, and CO-coated grains, outside the CO snow-line (dotted curve), are contoured in white.  \emph{Right} - A simplified representation of the emission regions of the main simple molecules is shown in the top panel, while the main dust thermal and scattered light emission regions are highlighted in purple and yellow in the bottom panel. Alongside, }the wavelength range of the emission is reported. The axes show the approximate logarithmic  distance from the central star, both in the radial and vertical direction. 
      \label{fig:pp7disk}
\end{figure*}

Since \enquote{Protostars and Planets VI}, the field has seen significant improvements in observational capabilities, i.e., the Atacama Large Millimeter/submillimeter Array (ALMA) becoming operational and the arrival of a new generation of high-contrast optical and IR instruments. These facilities have transformed the field of planet formation and have enabled both moderate resolution ($\sim$ 0.1" - 0.4", tens of au in nearby SFRs) statistical disk surveys and high resolution (down to $\sim$ 20 mas, $\sim$ 1 au in the closest disks) imaging studies of disks at radio and scattered light wavelengths. Recent observational results, combined with advances in disk physical and chemical modelling, have resulted in a much more detailed and complex picture of the planet forming environment. As illustrated in Figure~\ref{fig:pp7disk} and described in the subsequent sections, the planet forming materials, their physical conditions, and the observational tools required to trace them vary {\em greatly} as a function of position. The variety of physical environments present -- even within a single disk -- may shed important light on the known diversity of exoplanets. Nevertheless, improvements in data quality and sample size have possibly opened more questions than have been answered.

Outstanding questions include how accurately we know the dust and gas \textit{mass budget} of disks, how the mass is \textit{spatially distributed} and what its radial extent is,  how \textit{geometrically thick} and layered disks are and how this is connected to disk \textit{viscosity} and \textit{temperature}. Each one of these properties is of interest as it sets the stage for disk evolution, informs which processes are at play, and feeds back into corresponding planet formation theories. While targeted observations of disks have revealed -- in many aspects -- an increasingly diverse population, this chapter seeks to {\em unify} our understanding of the more general properties that regulate planet formation.

In this chapter, we will review our understanding of the fundamental properties of disks including the relevant observational techniques, molecular probes of physical processes, modelling methods, and each of these tool's respective caveats. The chapter is structured as follows. Sec. \ref{sec:fund_prop} describes our current understanding of each fundamental disk property.  Sec. \ref{sec:synthesys} discusses open questions and their implications for planet formation. Sec. \ref{Sec:summary} synthesizes our current picture of bulk disk properties, and summarizes key areas of improvement needed going forward. We note that many disk properties, their measurement, and the implications of such measurements are highly interdependent. We therefore decided not to separate the discussion as it has been done traditionally, for example presenting first a description of the dust in disks and then describing their gaseous component. For this reason some topics (e.g., dust opacity, disk thermal structure), that are relevant for different disk properties, are discussed in different parts of the text as needed.

\section{\textbf{FUNDAMENTAL DISK PROPERTIES}}
\label{sec:fund_prop}
\subsection{\textbf{Disk Mass}}

\subsubsection{\textbf{Dust mass: dust grain sizes, opacity and optical depth}}
\label{sec:dust_mass}

The total amount of observable solid material - which we will call dust - and the sizes of these  dust particles are among the two most crucial pieces of information to inform theoretical models of planet formation. The former tells us how much material there is (or remains) to form terrestrial planets or cores of giant planets. The latter, i.e., the sizes of dust particles, determines their aerodynamic behavior: how they interact with the gas and how they accumulate to eventually assemble the building blocks of planets. However, measuring  dust surface density and particle sizes is not straight-forward. Both of these quantities are generally entangled with the disk physical structure and the particle composition, impacting  their optical properties. Interpreting observations of the continuum flux of disks therefore relies on modeling and strong assumptions to derive the desired physical quantities. On the most basic level, this requires translating the amount of material and its optical properties to the emitted continuum intensity and vice versa.

\paragraph{Continuum emission from dust particles --} Neglecting scattering for now (see \autoref{Sec:scattering}), the outgoing intensity of a plane-parallel layer with homogeneous temperature and opacity can be calculated as
\begin{equation}
      I_\nu = B_\nu(\Td) \left(1 - e^{-\tau_\nu}\right),
\end{equation}
where $B_\nu$ is the Planck spectrum at the dust temperature, $\Td$, and $\tau_\nu$ the optical depth.

If the dust emission is optically thick ($\tau_\nu\gg 1$) then $I_\nu = B_\nu(\Td)$, and the dust temperature can be measured from the observed intensity $I_\nu$. For optically thin emission ($\tau_\nu \ll 1$), the outgoing intensity becomes
\begin{equation}
      I_\nu = B_\nu(T) \, \tau_\nu = B_\nu(T) \, \Sigd\,\kapabs.
      \label{eq:thin_emission}
\end{equation}
It can be seen that there are two factors that contribute to the frequency dependence of Eq. \ref{eq:thin_emission}: the Planck spectrum, which at (sub-)millimeter wavelength is approximately in the Rayleigh-Jeans limit, $\propto \nu^2$, and the opacity, which is a pure material property and commonly written as a power-law $\kapabs \propto \nu^\beta$ (although $\beta$ will itself be wavelength dependent in general).

These considerations already reveal some possible pathways to measure the desired physical quantities: in the optically thick limit, the intensity and temperature are linearly related. In the optically thin limit, the spectral index $\alpha$ defined as $I_\nu\propto \nu^\alpha$ is related to the material property as $\alpha = \beta +2$ (where the value 2 applies only in the Rayleigh-Jeans limit and might be smaller closer to the peak of the Planck spectrum), relating the spectral dependency of the emission to the emitting material.

From \autoref{eq:thin_emission} it can be seen that the dust surface density, the temperature profile or the opacity could be measured at any place in the disk if the other two quantities were known, which unfortunately is not the case for astrophysical sources.
Disentangling the right hand side of \autoref{eq:thin_emission} is, in principle, possible by using 1) a given or a parameterized model of the opacity, 2) the same for the temperature, and 3) enough wavelength coverage to constrain all parameters in these models \citep[e.g.,][]{CarrascoGonzalez2019}. We will discuss resulting constraints on the particle sizes in later sections (\autoref{Sec:scattering}) and focus in this section on the dust mass measurements and its caveats.

\paragraph{Continuum flux to dust mass conversion --}If there was a known appropriate \enquote{average} temperature of the disk $\bar T_\mathrm{d}$ \textit{and} an \enquote{average} dust opacity $\bar \kappa$, \textit{and} if the emission were optically thin, then the total flux could be used to infer a (rough) estimate for the disk mass since
\begin{align}
      F_\nu & = \frac{1}{d^2} \int 2\pi\,r\,I_\nu(r)\,\mathrm{d}r \nonumber                                  \\
            & = \frac{1}{d^2} B_\nu(\bar T_\mathrm{d}) \, \bar \kappa \, \int 2 \pi r \Sigd(r) \,\mathrm{d}r
      \label{eq:cont_flux_mass}
      \\
            & = \frac{B_\nu(\bar T_\mathrm{d}) \, \bar \kappa}{d^2} \, \Md, \nonumber
\end{align}
where $d$ is the distance to the source, $\Md = \int 2\pi\,r\,\Sigd \, \mathrm{d}r$ is the total dust mass of the disk and the integrals are over the entire disk. While obviously very approximate, this flux-to-mass conversion first proposed by \citet{Hildebrand1983} \citep[see also][]{Beckwith1986,Beckwith1990} has since then been used and discussed countless times in the literature, see \citet{D'Alessio1999} and \citet{D'Alessio2001} for pioneering work on the radiative transfer, \citet{Andrews2005} for early works on disk populations as well as recent discussions in \citet{Hendler2017} and \citet{Ballering2019}. At a first glance one might think that \autoref{eq:cont_flux_mass} is a fair relative measure of mass, but this assumption can fail if the compared disks are on average hotter or larger or possess significantly different grain properties.

\begin{figure}[htb]
      \includegraphics[width=0.95\hsize, trim= .6cm .1cm 1.4cm 1.1cm, clip]{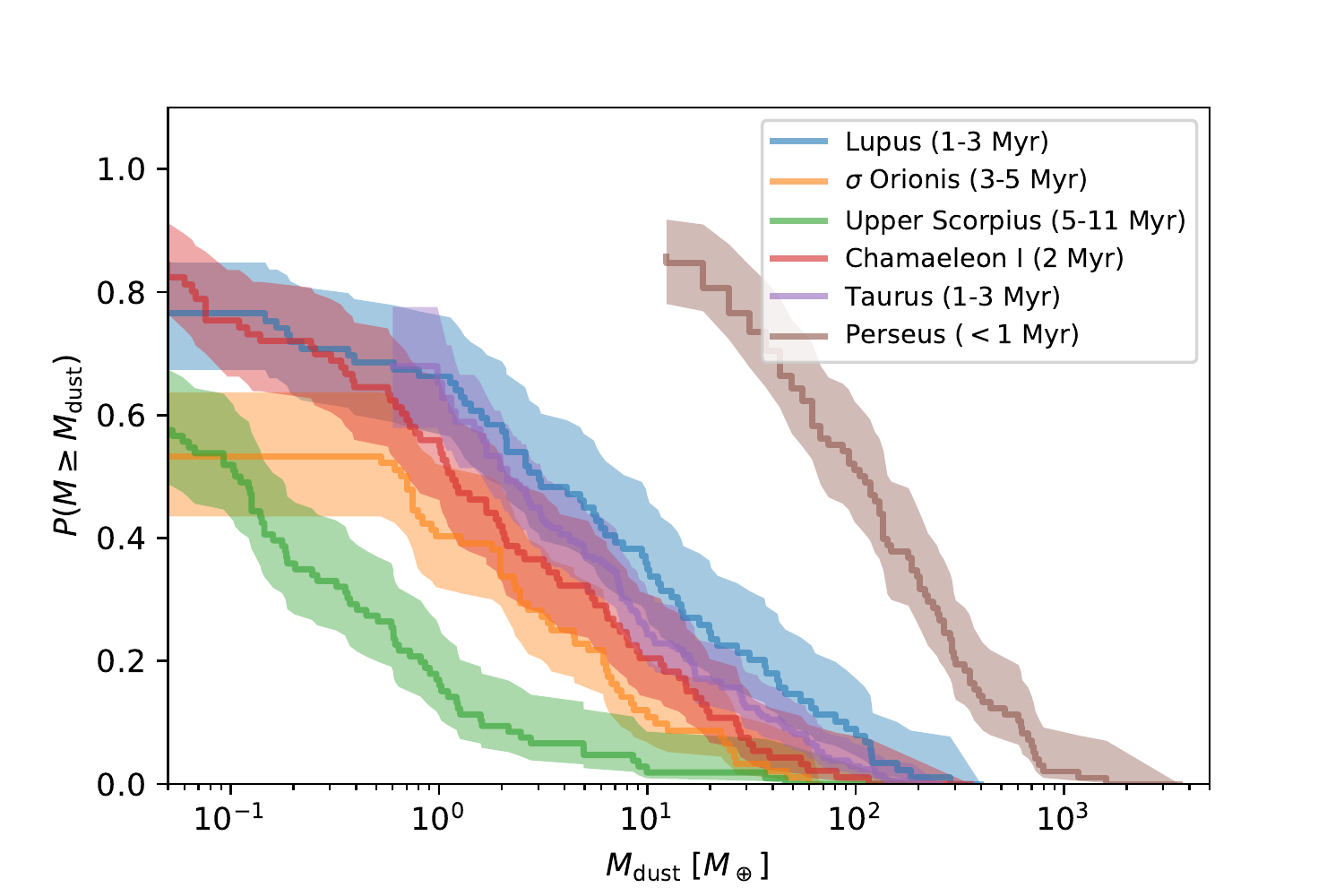}
      \caption{Cumulative disk mass distributions after \cite{Tychoniec2018} for different SFRs. Note: distributions do not reach unity because of the $\gtrsim 20\%$ of non-detections in the respective sample. Also, separating the disk and envelope contributions in Class 0/I sources is not trivial and may be a source of uncertainty.}
      \label{fig:tychoniec}
\end{figure}

With these caveats in mind, dust masses can now be derived for a large number of masses in several star forming regions. This was done, among other Star Forming Regions (SFRs), for
Taurus \citep{Andrews2013},
Lupus \citep{Ansdell2016},
Upper Scorpius \citep{Barenfeld2016},
Chamaeleon I \citep{Pascucci2016},
$\sigma$ Orionis \citep{Ansdell2018},
and some Class 0/I source in Perseus \citep{Tychoniec2018} (see PPVII Chapter by Manara et al., for a more detailed discussion). The cumulative mass dust distributions, derived from the observed fluxes and assuming a dust opacity of
$\kapabs = \SI{3.5}{cm^2/g}\times(\SI{870}{\micro m}/\lambda)$
\citep{Beckwith1990} are shown in \autoref{fig:tychoniec}. While some of the differences might stem from the caveats discussed above, significant differences are seen in the oldest disks \citep[Upper Sco, \SIrange{5}{11}{Myr},][]{Barenfeld2016} that show a factor of a few lower median dust masses compared to most other star forming regions. Even more significantly different are the youngest disks \citep{Tychoniec2018} which have more than an order of magnitude larger dust masses. A possible caveat to this discussion lies in the uncertainties on the determination of the timescales, i.e., of the stellar ages \citep[see e.g., ][]{Simon2019,Mullan2020,Pegues2021}. A more detailed discussion on this topic can be found in the PPVII Chapter by Manara et al.

This variation in mass for the different regions is suggestive that a large fraction of the initial dust reservoir is quickly processed. The dust may be aggregated into planetesimals or lost via processes like radial drift, whereby it is either accreted onto the star or hidden in high-optical depth regions. The radial speed of particles in the drift limit is $v_r = \epsilon\,\Vk$ \citep{Birnstiel2012} where $\epsilon$ is the dust-to-gas mass ratio and \Vk the Keplerian speed. This is broadly in agreement with these observations where the initial dust mass is draining quickly, but with decreased amounts of dust present (i.e., a reduced $\epsilon$), the time scale for dust depletion becomes longer. However, a detailed comparison between observations and populations of models has yet to be done.

\begin{figure}
      \includegraphics[width=\hsize]{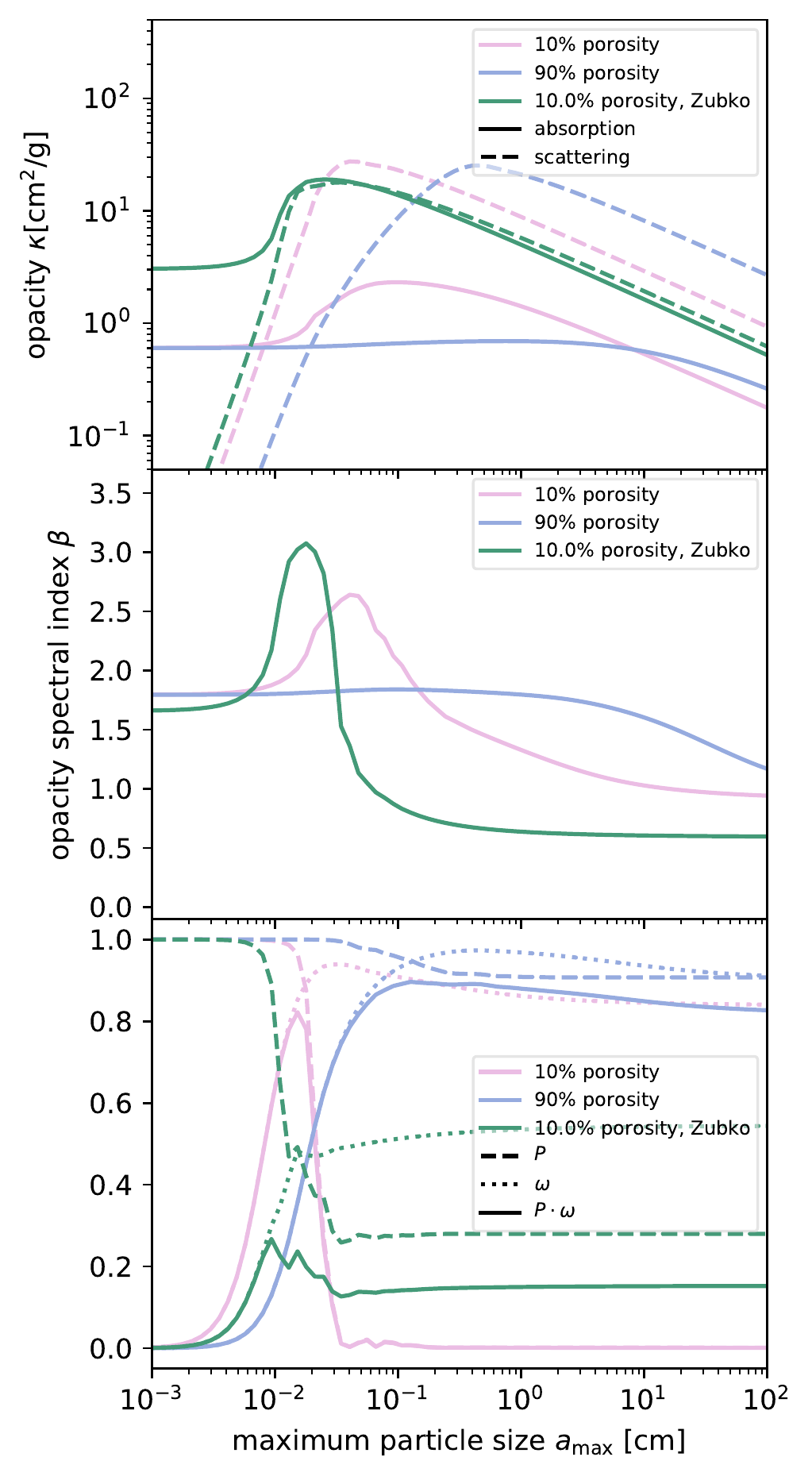}
      \caption{Results of Mie opacity calculations for distributions of particles of high (90\%) and low (10\%) porosity. The particle size distributions are always following the \citet{Mathis1977} power law of $n(a)\propto a^{-3.5}$ and extend from a minimum size of \SI{0.1}{\micro m} to a maximum particle size \amax. \emph{Top.} Absorption (solid line) and scattering mass opacities (dashed). \emph{Middle.} Opacity spectral index $\beta$ for both porosities. \emph{Bottom.} The polarization fraction of 90$^\circ$ scattering, $P$, the albedo $\omega$, and their product.}
      \label{fig:opacities}
\end{figure}

\paragraph{Changing paradigms --}Several of the above mentioned assumptions have recently been questioned. While it was expected that the inner disk regions could be optically thick at radio frequencies, the contribution of optically thick emission was often thought to be small \citep[see][]{Ricci2012}. \citet{Tripathi2017} and \citet{Andrews2018a}, however, found that the disk luminosity scales with the disk area. This fact is not straight-forwardly explained for optically thin disks (although it is an expected outcome for drift-limited particles, see \citealp{Rosotti2019}). \citet{Tripathi2017} and \citet{Andrews2018a} suggest therefore, that optically thick emission spanning $\sim$30\% of the disk area would predict the same scaling of disk size and luminosity. Subsequent observations have shown that optically thick sub-structure is ubiquitous in bright disks \citep[e.g.,][]{Andrews2018, Huang2018b, Long2018} which might mean that indeed a substantial fraction of the emission could stem from optically thick regions (see also the discussion of \citealp{Tazzari2021}). If this were the case, the disk mass budget in solids may be underestimated. The problem is even greater if scattering opacity is considered \citep[e.g.][]{Zhu2019,Liu2019}.

Beyond the optical depth, the dust optical properties are a further source of uncertainty. This encompasses the unknown abundances and mineralogy of the particles as well as the grain structure (porosity and fractal dimension). As an example, \autoref{fig:opacities} shows Mie opacities at $\lambda=\SI{1}{mm}$ for a particle size distribution up to a grain radius of \amax for particles with a composition similar to \citet{Birnstiel2018}, but including 10\% and 90\% porosity. The figure shows several key features: the size averaged opacities (top panel) have a characteristic shape with a constant opacity for sizes much smaller than the wavelength and a decreasing opacity for grain sizes much larger. Between those regimes, around $2\pi\,\amax=\lambda$, low-porosity grains have a distinct peak. As discussed in \citet{Kataoka2014}, this regime change happens at particle sizes where $2 \pi\amax \,f = \lambda$ with $f$ being the filling factor of the particle. Furthermore, the interference peak becomes smoothed out and has effectively vanished for a porosity of 90\%.

The center panel in \autoref{fig:opacities} shows the opacity spectral index $\beta$, which reaches its lowest values for the largest maximum particle sizes \amax. Again, the porosity and material composition can substantially affect these curves' absolute values.

The bottom panel in \autoref{fig:opacities} depicts the single scattering albedo $\omega$ and the 90$^\circ$ scattering probability $P$. It can be seen that particles smaller than $\lambda/(2\,\pi)$ have a low albedo while large particles have a high albedo. While this seems like large particles are effective at scattering, it turns out that they are mainly forward scattering. The product $P\cdot \omega$ is therefore peaked at $\amax = \lambda/(2\pi)$ for most compositions and the amount of scattering in the peak and for larger sizes strongly depends on the particle composition and porosity.

In all plots, we included a material mix that is identical to the case labeled with ``10\%'', but have replaced the organics (using optical constants from \citealp{Henning1996}, similar to \citealp{Pollack1994}) with the carbonaceous material from \citet{Zubko1996}. The result is a strong difference causing much higher scattering \emph{and} absorption opacities while reducing the albedo.

These results already indicate how strongly the appearance of protoplanetary disks and their observational interpretation hinges on the microphysical (optical) properties of dust. This point is especially salient, since this discussion does not even consider more realistic opacity models (see \autoref{sec:obs_settling}) or the temperature dependence of the optical constants \citep[e.g.][]{Boudet2005,Coupeaud2011,Demyk2017a,Demyk2017b,Birnstiel2019}.

\begin{summarybox}
      In summary, while significant assumptions are made to derive dust masses from continuum fluxes, there appears to be a mass loss of two orders of magnitude from young to old disks (see Fig. \ref{fig:tychoniec}) -- a difference that is difficult to attribute to just uncertainties of the opacities, or the optical depths. In the future, it should be quantified how differences in the grain properties, the optical depths, or the radial temperature profiles affect these conclusions.
\end{summarybox}

\subsubsection{\textbf{Gas mass}}
\label{sec:gas_mass}

Determining protoplanetary disk gas mass, commonly assumed to be the bulk of the disk mass, is one of the most significant outstanding questions in the field of star and planet formation. Disk mass determines disk physics and evolution, from disk formation all the way through planet formation. Although it is essential to constrain disk gas masses, measuring them is anything but simple \citep[see e.g.,][]{Bergin2018}, as we will explain in this section.

Molecular hydrogen (H$_2$) is the main gaseous species present in disks, but its emission is faint due to the details of its molecular physics \citep{Field1966}.
The large energy spacing of its fundamental ground state transition, combined with the weaker quadrupole Einstein A coefficient, makes H$_2$ emission extremely faint in cold environments - such as the outer regions of protoplanetary disks - with typical gas temperatures $T_\mathrm{gas} = \SIrange{20}{30}{K}$. Only at much higher temperatures, $T_\mathrm{gas} > 100$ K, very close to the central star, does H$_2$ emission become relevant, but it is insensitive to the bulk disk mass \citep{Thi2001,Pascucci2006,Carmona2008,Bitner2008,Bary2008,Pascucci2013}. Given that direct detection of the cold molecular hydrogen is extremely difficult, one must rely on indirect gas mass tracers.\\

\textbf{Gas mass from HD emission - }The closest molecule to H$_2$ is its less abundant isotopologue hydrogen deuteride (HD). HD chemistry is similar to that of H$_2$ as it does not freeze-out onto grains. In fact other molecules, e.g., CO and less volatile species, cannot survive in the gas phase at low temperatures but stick onto the icy grains through freeze-out. Also HD, like H$_2$, can shield itself from the photodissociating UV photons, but at reduced efficiency \citep{Wolcott-Green2011}.
The abundance of HD with respect to H$_2$ is $\sim 3 \times 10^{-5}$, obtained assuming the local [D]/[H] value \citep{Prodanovic2010}. In contrast to  H$_2$, HD has a small dipole moment which allows dipole transitions ($\Delta J = 1$). The energy needed to excite the HD fundamental rotational level ($J=1-0$) is 128 K and this means that at a temperature \Tg between $20$ and $100$ K the expected emission of HD is much larger than that of H$_2$. Although this line does not directly trace the bulk of the cold H$_2$ gas, its emission can be used to measure the bulk gas mass relying on physical-chemical models of the disk structure \citep[see e.g.,][]{Bergin2013,Kama2016,Trapman2017}.   The fundamental rotational transition of HD is at 112 $\mu$m and it was covered with the PACS instrument on the \emph{Herschel Space Observatory}. This transition has been targeted for a sample of close and bright protoplanetary disks, for a total of only 3 detections: TW Hya \citep{Bergin2013}, DM Tau, and GM Aur \citep{McClure2016}. The clear detection of HD in TW Hya was an important result as an unexpectedly high disk mass of 0.06 $M_{\odot}$ was determined for a relatively old disk  of $\sim$ 10 Myr (see Fig. \ref{Fig:TWHya} and Sec. \ref{ex_TWHya}). However, there are some caveats in the conversion of HD mass into total disk mass. The first complication is that the emitting layer of the HD 112 $\mu$m line is elevated above the midplane, where the gas temperature is larger than 30 K. This implies that a good knowledge of the disk vertical structure is needed in order to correctly estimate the disk mass \citep{Trapman2017,Calahan2021}. The second, more practical, problem is that there is no current or planned facility that covers the HD $J=1-0$ transition to carry out a large unbiased sample of disks \citep{Kamp2021}.\\

\textbf{Gas mass from CO emission - }Carbon monoxide (CO) and its less abundant isotopologues are often used as tracers of gas properties, structure and kinematics in disks. CO is the second most abundant molecule after H$_2$ and it is the main gas-phase carrier of interstellar carbon. Furthermore, CO is chemically stable, it has a well studied and relatively simple interstellar chemistry, which is readily implemented in physical-chemical models at different levels of sophistication \citep[e.g.,][]{Bruderer2012,Woitke2011,Miotello2014,WilliamsBest2014}. All of these factors make CO a promising gas mass tracer, as its distribution could be quite directly linked to that of molecular hydrogen without relaying excessively on assumptions about disk chemistry \citep[see e.g.,][]{Kamp2017}.

\cite{Aikawa2002} provided a very simple description of the CO distribution in disks: CO should be abundant in a warm
molecular layer defined by two boundaries. The lower boundary is set by CO freeze-out in the cold disk midplane and the upper is set by photo-dissociation from UV photons coming from the central star or an external radiation field. The first process reduces the amount of gas-phase CO, which is frozen onto dust grains at 20~K, for a pure CO ice with a binding energy of 855~K measured in the laboratory \citep{Bisschop2006}. This can, however, vary between 17~K and 30~K varying the assumed density and binding energy in mixed ices \citep[see][and reference therein]{cleeves14,Harsono2015}. The latter, CO photodissociation, is controlled by line processes that are initiated by discrete absorption of photons into predissociative excited states and is thus subject to self-shielding \citep{BallyLanger1982,vanDishoeck88,Viala1988}. An additional process that may be important in the outer disk regions is photodesorption of CO molecules from the ices, as shown for the AS 209 disk \citep{huang16} and IM Lup \citep{Pinte2018}.

Above all, CO is a very practical mass tracer with its relatively high abundance and  detectability at (sub-)mm wavelengths. The main isotopologue of carbon monoxide, $^{12}$CO, is so abundant in disks that its emission is mostly optically thick and mainly provides information about the disk temperature and kinematics. Nevertheless, $^{12}$CO emission has been used in earlier studies to determine the cold gas mass, using simple formulae, and with large uncertainties due to the enhanced optical depth of the lines \citep{Thi2001}. In order to probe column densities, instead,  one needs to employ more optically thin tracers, such as low-lying rotation emission of $^{13}$CO and C$^{18}$O. Similarly to $^{12}$CO, less abundant isotopologues are prone to mutual shielding \citep{Visser2009} that selectively affects their abundance.  Isotope-selective processes, as for example isotope-selective photodissociation, were proven to be important when interpreting $^{13}$CO and C$^{18}$O emission for measuring disk masses. Mass could be underestimated by up to two orders of magnitudes if isotope selective processes are neglected. Such processes were implemented in physical-chemical models by \cite{Miotello2014} and more recently by \citep{Ruaud2021}, and \cite{Miotello2016} provided simulated CO fluxes as well as analytical prescriptions. Recently $^{13}$C$^{17}$O has been observed in a limited sample of bright disks and is a promising mass tracer, as it minimizes the uncertainty given by line optical depth \citep{Booth2019,Booth2020,Zhang2020}. Unfortunately, such observations are still very time consuming and are not practical for a large, unbiased sample of disks.

A crucial complication to measuring gas masses using CO isotopologue  lines comes from  uncertainty in the C/H ratio in disks. Observations of the HD fundamental line in TW Hya showed in fact that CO-based gas masses can be from one to two orders of magnitude smaller than HD-based disk masses, even considering isotope-selective processes and CO freeze-out in the analysis \citep[e.g.,][]{Bergin2013,Favre2013,Cleeves2015,Trapman2017,Calahan2021}.
This potential inconsistency can be  explained by locking up of carbon and oxygen rich volatiles as ice in larger bodies, leading to low observed CO fluxes. Interestingly enough, the CO-HD discrepancy is not there for warmer disks around Herbig stars \citep{Kama2020}, supporting the idea of volatile lock-up happening when freeze-out is more efficient. A similar trend between C depletion and disk temperature has been recently found with [CI] Atacama Compact Array (ACA) observations by \cite{Sturm2022}.

In the past few years, ALMA has also significantly enhanced the disk sample size by surveying continuum and CO isotopologue emission of disks in many different nearby SFRs: the $\sim1$ Myr-old to $\sim10$ Myr-old Lupus, Chamaeleon I, Orion Nebula Cluster, Ophiuchus, IC348, Taurus, and Corona Australis, $\sigma$-Orionis, $\lambda$-Orionis, and Upper Scorpius regions \citep{Ansdell2016,Pascucci2016,Eisner2016,Cieza2019,Long2018,Cazzoletti2019,Ansdell2017,Barenfeld2016,Ansdell2020}. One of the main results of these surveys, unfortunately carried out with short integration times, is that CO isotopologue emission is fainter than expected, based on pre-ALMA observations  of large, bright disks, leading to inferences of low dust and gas masses \citep{Ansdell2016,Pascucci2016,Long2017,Miotello2017,Manara2018}. These observations can be interpreted as a sign of fast gas dispersal or of quick chemical evolution. The second hypothesis, similar to the TW Hya scenario, is supported by the large observed mass accretion rates, which would not be sustainable in the first scenario \citep{Manara2016}. Observations of other molecular species such as C$_2$H and N$_2$H$^+$ also supports this scenario \citep[see e.g.,][and Sec. \ref{Mdisk_new} for more detail]{Bergin2013,Cleeves2018,Miotello2019}.

\begin{figure*}[ht]
      \centering
      \includegraphics[width=0.75\textwidth]{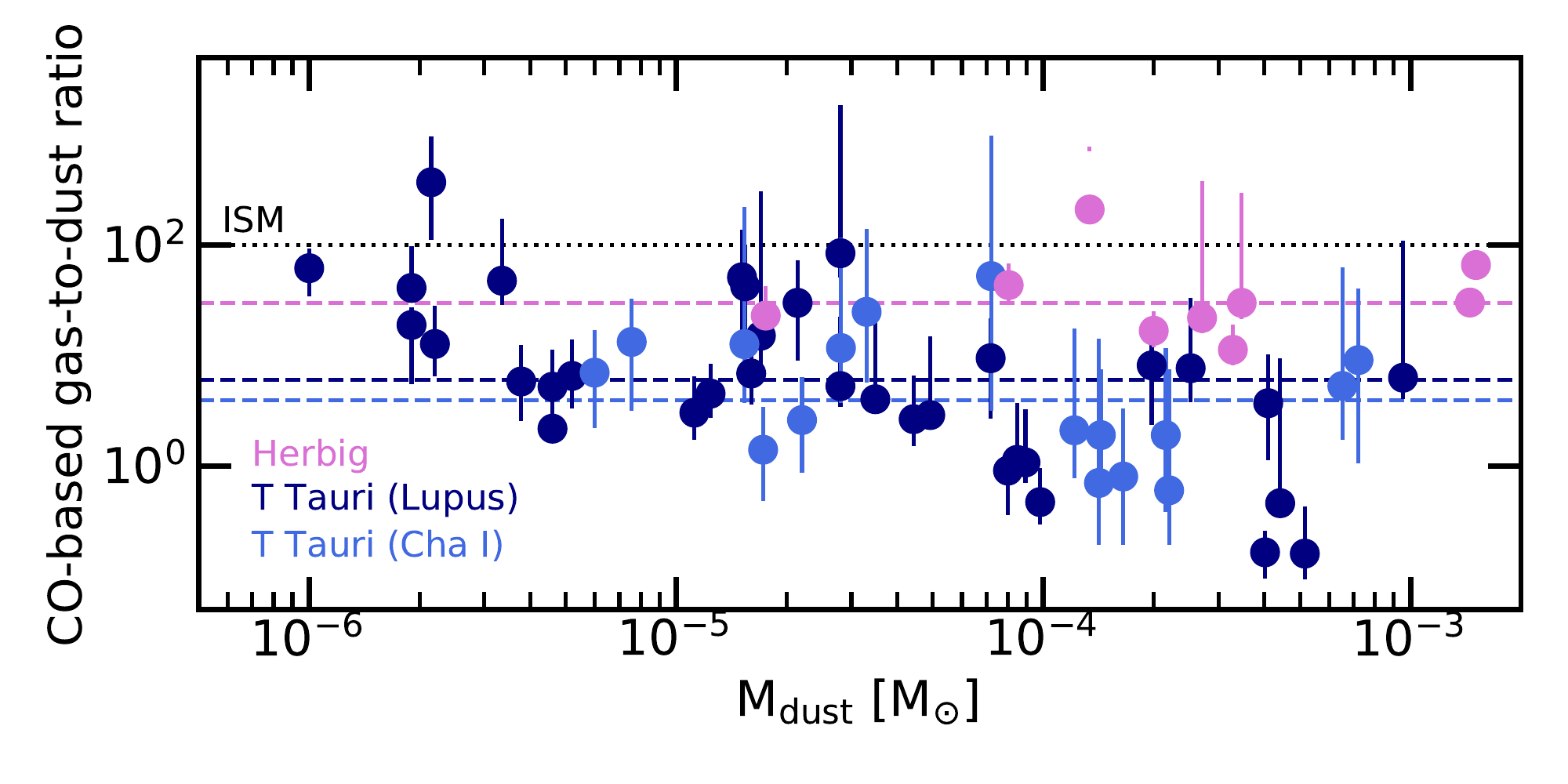}
      \vspace{-6mm}
      \caption{CO-based gas mass measurements obtained from DALI models \citep{Miotello2016} as a function of dust mass. Disks around T Tauri stars from the Lupus SFR are shown in dark blue and from the Chamaeleon I SFR in light blue \citep{Miotello2017,Long2017}, while disks around Herbig stars from the literature \citep{Miley2019,Taun2019} are shown in pink.}
      \label{Fig:g2d}
\end{figure*}

\cite{Kama2020} recently published \textit{Herschel} HD upper limits for 15 individual Herbig disks, that allowed them to set upper limits to their masses. Nearly all the disks are constrained to $M_\mathrm{gas} \leq 0.1 M_\odot$, ruling out global gravitational instability. A strong constraint is obtained for the HD 163296 disk mass, $M_\mathrm{gas} \leq 0.067 M_\odot$, implying a gas-to-dust ratio $\leq$ 100. This HD-based mass limit is towards the low end of CO-based mass estimates for the disk, highlighting that gas-phase CO depletion in HD 163296 may not be as severe as for the T Tauri disks, at most a factor of a few. From chemical models it is clear that volatile carbon and oxygen lock-up can be better facilitated when a large fraction of the disk material is frozen-out onto grains, i.e., in colder environments and when vertical mixing is facilitated \citep{Kama2016,Yu2017}. Following this reasoning, at a zeroth level approximation, one could expect volatile carbon and oxygen depletion to be more severe in colder T Tauri like disks, rather than in warmer Herbig disks. In order to empirically test this hypotheses with observations, in Fig. \ref{Fig:g2d} CO-based gas-to-dust ratios of T-Tauri disks in Lupus and Chamaeleon I, shown in dark and light blue respectively, are compared with gas-to-dust ratios of Herbig disks taken from the literature \citep{Miley2019,Taun2019}, shown in pink. The CO-based gas masses have all been determined using the DALI model results from \cite{Miotello2016}, as shown by \cite{Long2017} for Chamaeleon and by \cite{Miotello2017} for Lupus. Here we apply this method to a few of Herbig disks for the first time. The uncertainty on the disk mass, shown by the error bars, reflect the wide range of disk parameters (such as e.g., disk radius, slope of the surface density distribution, disk scale-height, and flaring) considered in the large gird of models by \cite{Miotello2016}. Some of the more extreme CO-based gas-to-dust ratios are found for edge-on disks and large cavity transition disks, which are likely not perfectly represented by the full-disk models that have been employed \citep[see Fig 6 of][]{Miotello2017}. The dashed colored lines in Fig. \ref{Fig:g2d} show the median of the derived CO-based gas-to-dust ratios for the three samples. While the medians obtained for T Tauri disks, which are similar for Lupus (6.01) and Chamaeleon (3.95) disks, are roughly one order of magnitude below the ISM gas-to-dust ratio of 100 (shown by the dotted black line), the median obtained for Herbig disks is of 30, only a factor of 3 lower that the ISM value. Fig. \ref{Fig:g2d} shows measurements for disks that have been detected in $^{13}$CO and the sample is limited in size. Deeper CO surveys of T Tauri disks and, most importantly, a uniform survey of Herbig disks would help refining this comparison \citep[see, e.g.,][]{Stapper2021}.  \\

\textbf{Gas mass from other atomic and molecular tracers - } The [OI] $^3P_1 - ^3P_2$ 63.2 $\mu$m line is exceptionally bright in protoplanetary
disks. It was detected in the majority of the T Tauri disks observed by the \emph{Herschel} open time key program GASPS
\citep{Dent2013}. Given its high excitation energy (227 K),
the [OI] 63.2 $\mu$m line arises from the surface layers of the disk and is more sensitive to temperature variations than total mass variations. Nevertheless, the [OI] 63.2$\mu$m /CO $(J=2-1)$ line ratio was found to be a good diagnostic for breaking disk modeling degeneracies. Theoretical models find it to be useful for obtaining disk gas mass estimates within one order of magnitude, purely from gas observations \citep{Kamp2010,Kamp2011}. In general, combined studies of different emission lines from different tracers are a very effective way to constrain disk masses in a more robust way, as shown below in the ase of \autoref{ex_TWHya}.

Observations of multiple rotational transitions from a single molecule, e.g., carbon monosulfide (CS) observed in TW Hya by \cite{Teague2018}, can be used to place meaningful limits of the local H$_2$ density using excitation.
For example, \cite{Teague2018} analyzed three transitions of CS ($J=7-6$, $J=5-4$, and $J=3-2$), and assuming local thermodynamic equilibrium (LTE) so that $T_\mathrm{ex} =
      T_\mathrm{kin}$, they measure a lower limit to the gas temperature where the lines are emitted. In the optically thin limit, the integrated flux $W=\int T_\mathrm{B} dV$, can be linked to the level population of the upper energy state, specifically the column density of the upper energy level $N_u$:
\begin{equation}
      \frac{N_u}{g_u}=\frac{\gamma_u W}{g_u},
\end{equation}
where $g_u$ is the degeneracy. In the LTE assumption, one can relate:
\begin{equation}
      \ln \left( \frac{\gamma_u W}{g_u}\right)= \ln N - \ln Q(T) - \ln C_{\tau} - \frac{E_u}{k T_\mathrm{ex}},
\end{equation}
where $N$ is the CS total column density, $Q$ is the partition function, and $C_{\tau}$ is the optical depth correction. Finally relating $N_u$ to $N$ through:
\begin{equation}
      N_u = \frac{g_u N}{Q(T)}\exp \left( \frac{-E_u}{k T_\mathrm{ex}}\right),
\end{equation}
allows one to fit for $T_\mathrm{ex}$ and $N$(CS) self-consistently and accounting for possible optical depth effects. \cite{Teague2018} then perform a non-LTE analysis, fitting the spectra with slab models, which allows them to constrain the local gas density $n_\mathrm{H_2}$, together with $N$(CS) and $T_\mathrm{kin}$. Using the minimum $n_\mathrm{H_2}$ they obtain the minimum $\Sigma_\mathrm{gas}$, which integrated give a lower limit to the total disk mass. Interestingly, the CS-based lower limit for the disk mass of TW Hya of $3\times 10^{-4} M_{\odot}$ is compatible with, albeit far lower than, the HD-based mass estimate by \cite{Bergin2013} (see Fig. \ref{Fig:TWHya}).

Recently, \cite{Trapman2022} have shown the potential of using a combination of N$_2$H$^+$ and C$^{18}$O to reduce the uncertainty on the CO-based measurements of protoplanetary disk masses. Using disk structures obtained from the literature, the authors set up thermo-chemical models for three disks, and show that the N$_2$H$^+$(3-2)/C$^{18}$O(2-1) line ratio scales in fact with the disk CO-to-H$_2$ ratio, braking the degeneracy between gas mass and C abundance. The main caveat of this interesting novel approach is however the uncertainty on the cosmic ray ionization rate $\zeta_\mathrm{CR}$. \cite{Trapman2022} propose that further observations of ionization tracers such as H$^{13}$CO$^+$ and N$_2$D$^+$,
      that constrain the cosmic ray ionization rate in disks, are needed.\\

\textbf{Dynamical constraints on the gas mass  - } Ideally one would like to directly constrain disk mass, without relying on indirect tracers. This can be attempted by tracing the disk gravitational mass through dynamical studies. This idea was initially proposed by \cite{Rosenfeld2013}, who showed that the disk self-gravity can be linked to the orbital velocity. The latter can be determined by spatially and spectrally resolved gas observations: it diverges from pure Keplerian rotation if the emission layer of the gas tracer is elevated from the midplane and if a strong negative pressure gradient is present (the orbital velocity becomes sub-Keplerian), or when the gravitational potential is not dominated only by the central star but the disk mass contribution becomes relevant (the velocity becomes super-Keplerian). \cite{Veronesi2021} have modeled the CO rotation curve of the massive Elias 2-27 disk \citep{Paneque2021} with a theoretical rotation curve including both the disk self-gravity and the star contribution to the gravitational potential. Including disk self-gravity leads to a better match to the observation and allows them to obtain the first dynamical estimate of the disk mass, which is is 17\% of the star mass, meaning that it could be prone to gravitational instabilities. This method is promising, as it allows to constrain the gravitational mass without relying on the emission of a specific tracer. On the other hand, by definition, such measurements can only be carried out in massive disks, relative to their host stars, and require high spectral and spatial resolution, which requires long integration times. Although not proven observationally yet, gravitational instabilities may leave kinematic perturbations, known as the ``GI Wiggle'' \citep{Hall2020}. Recently, \cite{Terry2021}, have determined an approximately linear relationship between the amplitude of the wiggle and the host disk-to-star mass ratio, and therefore the disk mass. Such measurements should be possible with the spatial and spectral resolution provided by ALMA for the brightest and most massive disks.

\begin{figure}[t]
      \includegraphics[width=\textwidth]{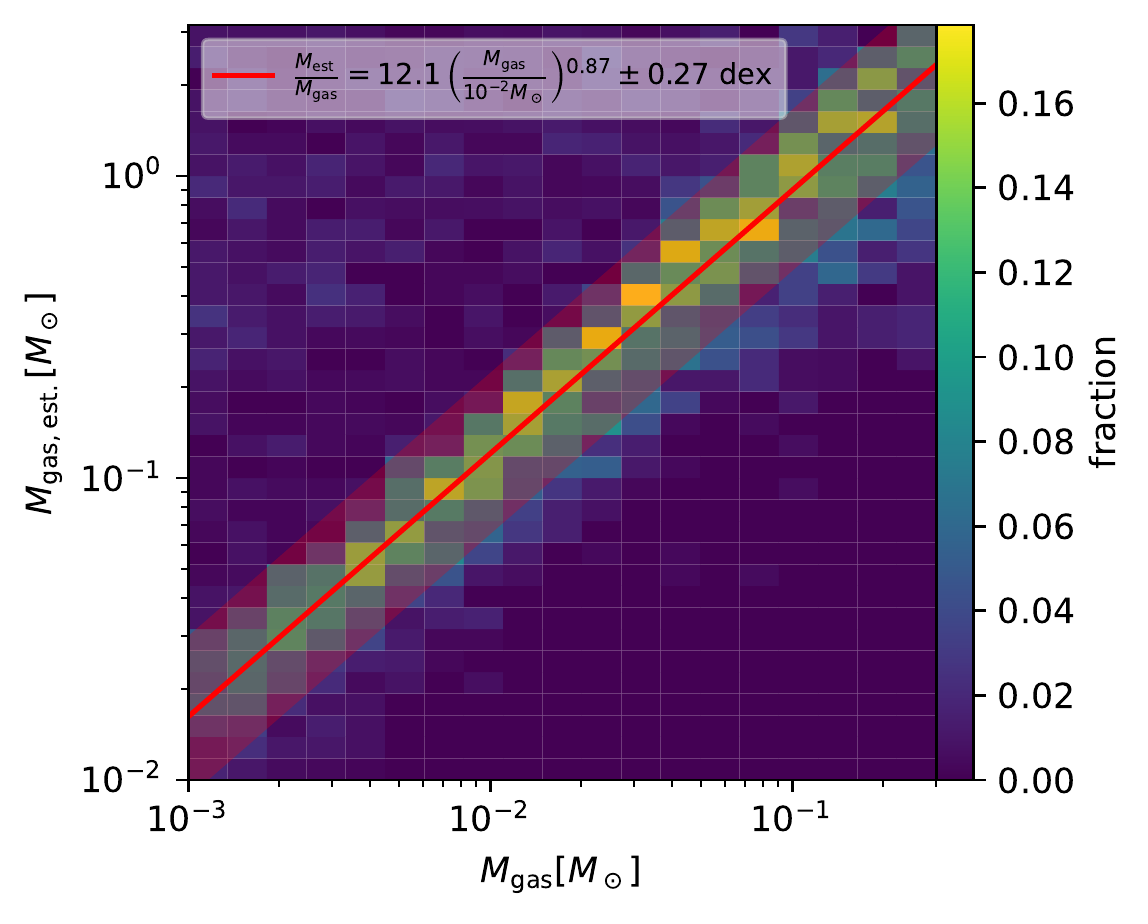}
      \caption{Gas masses derived with the \citet{Powell2017} approach for over 7000 disk models plotted versus the actual gas disk masses of the models, after \citet{Franceschi2022}. The plotted fraction corresponds to the number of simulations in that grid cell divided by the total number of simulations in this \Mg range. Overall, 44\% of the simulations are inside the shown correlation, but 20\% of the estimated gas masses are outside the range of the plot. The fitted correlation shows a close-to-linear scaling but a general over-estimation by an order of magnitude.}
      \label{fig:powell}
\end{figure}

\cite{Powell2019} have also proposed an alternative dynamical method for constraining disk masses, that relies on the location of the so-called dust lines, i.e., the apparent outer boundary of the continuum emission at different wavelengths. Dust continuum observations tend to find an anticorrelation between the observed wavelength and the disk radial extent \citep[e.g.,][]{Andrews2012,deGregorio-Monsalvo2013}. Models of dust growth and evolution generally produce a radial size sorting with larger grains inside \citep{Garaud2007,Brauer2008,Birnstiel2010,Okuzumi2012}. Approximating the dust thermal emission with \autoref{eq:thin_emission}, $I_\nu \propto B_\nu(\Td) \, \Sigd\, \kapabs$, we see that the Planck function and the dust surface density generally decrease with radius. With radially decreasing particle sizes, the opacity, however increases as long as particles have maximum sizes $2\,\pi \, \amax > \lambda$, but then steeply decreases to a constant values for smaller \amax (see \autoref{fig:opacities}, top panel). This steep decrease in the opacity causes a drop in the radially decreasing continuum emission, often leaving the outer emission below the typical noise level of observations. This truncation of the emission is not a truncation of the dust surface density, but instead a combination of the radial particle size sorting and the interference feature of Mie theory and therefore wavelength dependent. If the observed outer edge of the continuum emission is indeed linked to this opacity/size sorting feature and not instead caused by sub-structure in the dust surface density, then this radius can be used to constrain the particle size at this radial position in the disk. Further, assuming particle sizes are limited by radial drift, one can relate the particle size to the dust surface density \citep[see][Eq.~18]{Birnstiel2012}. This drift limit is derived from the assumption of the growth time scale and the drift time scale being equal, yielding $\amax \propto \Sigd$. If one additionally assumes that these time scales also need to match the disk age, one can put an additional constraint on the dust-to-gas ratio $\epsilon \sim (\Ok\,t_\mathrm{disk})$ \citep{Powell2017,Powell2019}. Thus, the measurement of the dust-line position becomes a constraint on the gas surface density at that position. It should be noted that equating all time scales is only a valid assumption in the outermost dust reservoir, as it is not sustained by dust influx from outside. This method was recently tested in \citet{Franceschi2022} who found indeed an almost linear correlation between the dust mass contained in their models and the mass derived with the method of \citet{Powell2017}, see \autoref{fig:powell}. However due to several coarse assumptions and the limitations outlined above, the derived masses derived in this way were generally overestimated by an order of magnitude. \\ 

\begin{summarybox}
      In summary, the total disk mass remains a fundamental, yet elusive, quantity. All methods described above have their flaws and rely on different types of assumptions. This needs to be taken into account when using disk mass measurements, for example in disk population studies. More implications and future perspective on the topic of disk masses are discussed in Sec. \ref{Mdisk_new}.
\end{summarybox}

\subsection{\textbf{Disk radial structure: $\Sigma(R)$, $R_\mathrm{out}$, and $T(R)$}}

The radial distribution of disk mass determines where planets can form and with what mass, at least initially. Over the disk lifetime, the distribution of mass is expected to evolve significantly. How this proceeds in detail is still an open question.  
Disk evolution leads to mass accretion onto the central star \citep{Hartmann2016}, but part of the disk content can be dissipated by high-energy radiation driven winds or magnetic torques \citep{Alexander2014,Gorti2016} and external processes such as stellar encounters \citep[e.g.,][]{Clarke1991,Pfalzner2005} and external
photoevaporation \citep[e.g.,][]{Clarke2007,Facchini2016,Winter2020}. Depending on which process is dominant, the disk radial structure will be considerably different. In particular this will impact the surface density distribution, $\Sigma(R)$, and the disk outer radius, $R_\mathrm{out}$, two parameters that determine the architecture of the forming planetary systems \citep{Morbidelli2016}. Reliable observational measurements of $\Sigma(R)$ and $R_\mathrm{out}$ would therefore be key, but there is no consensus yet either on their physical or empirical definition, as well as on the best tracers and observational approaches.

\subsubsection{\textbf{Dust Surface Density (\Sigd): from modeling of sub-mm continuum emission}}

\label{sec:sig_d}

As outlined in \autoref{sec:dust_mass}, the emitted intensity at a given position of the disk is at the very least (neglecting scattering) a line-of-sight integral over the opacity, temperature, and dust density. The low vertical extent of the dust disk along with the low opacities at (sub-)mm wavelength still make \autoref{eq:thin_emission} in many cases a decent approximation for large parts of the disk. For a fixed value of the dust opacity, and an assumed or modeled temperature profile, the surface density can be directly constrained from the observed intensity profile. \citet{Andrews2009,Andrews2010}, for example, simultaneously modeled the visibilities \& SED with a self-similar profile \citep{Lynden-Bell1974} and found a wide range in disk sizes and masses that overall do agree with the surface densities implied by the Minimum Mass Solar Nebula  (MMSN) \citep{Weidenschilling1977b,Hayashi1981}, although not the shape. Comparing the disks in the Lupus star forming region to the MMSN, \citet{Tazzari2017} found that observed disks have solid surface densities that are larger and follow a less steep radial surface density profile. The determined masses of the Lupus disks were found to be overall less massive than the MMSN.

\citet{Isella2010}, \cite{ Guilloteau2011}, \citet{Perez2012}, \citet{Tazzari2016} and others dropped the assumption of constant particle size and opacity and generally found that the data was best modeled with a radially decreasing particle size, in broad agreement with models of dust evolution \citep[e.g.][]{Birnstiel2012}. Modeling the dust surface density with a self-similar profile \citep{Lynden-Bell1974} and a radially varying particle size \citet{Tazzari2016} found a best matching radial exponent parameter of $\gamma \sim 1$.

Until recently, high optical depth was thought to play only a minor role since the dust masses required to produce the observed fluxes, distributed over the emitting area of the disk would yield low optical depth everywhere apart from inside $\simeq \SI{10}{au}$ \citep[e.g.][]{Ricci2010b}. \cite{Pietu2014} proposed then that compact disks could be potentially massive, hiding material in optically thick dust emission. More recently ALMA Long Baseline observations have iconically changed our view of the mass distribution in disks \citep[e.g.][]{ALMAPartnership2015,Andrews2016,Andrews2018,Huang2018b}, with mostly axisymmetric ring structures dominating the emission in many disks.

Well resolved observations at multiple wavelengths allow to partially disentangle the degeneracy between temperature, dust surface density, and dust opacity since  temperature and surface density are only functions of radius (for the vertically isothermal mid-plane) while the wavelength dependence at a given radius is only given by the Planck function and the opacity. This technique has been used in, for example, \citet{CarrascoGonzalez2019}, who find agreement with expectations of size-dependent dust trapping in pressure maxima.  This underlines the fact that the particle sizes are not constant throughout the disk as a constant opacity value would imply.

As can be seen in \autoref{fig:opacities}, scattering cannot be ignored as the scattering opacity -- depending on the grain composition -- can match or exceed the absorption opacity for large \amax. Also for scattering an analytical solution can be found in the plane-parallel two-stream approximation, following \citet{Rybicki1979}. This solution has been derived in \citet{Miyake1993, Birnstiel2018}\footnote{Both solutions mostly follow the notation of \citet{Rybicki1979} and are mathematically equivalent despite slightly different notation.}. Modeling sub-structured disks with scattering \citep[e.g.,][]{Sierra2021,CarrascoGonzalez2019} yields tight constraints on the dust surface densities, though only within the framework of the assumed opacity model. These models also require large (millimeter to centimeter sized) grains, which is inconsistent with results derived from polarization \citep[e.g.][]{Lin2020}, see \autoref{sec:dust_properties}. The data also show good agreement with models of dust traps since the \textit{outer} dust ring in the disk of HD~163296 (found to be a dust trap in \citealp{Dullemond2018}) was confirmed to harbor a local increase in the dust surface density and in the grain size \citep{Sierra2021}. For the \textit{inner ring}, however \citet{Dullemond2018} did not find a significant radial confinement of the dust (i.e. no trapping is required and the dust surface density might just follow a gas surface density increas at constant dust-to-gas ratio). This agrees with \citet{Sierra2021} who also did not find evidence for a grain size maximum at the inner ring location.

\begin{summarybox}
      In summary, measurements of the radial dust surface density leave us with no clear picture yet. While lower resolution studies found disks often to be less massive than the MMSN, high-resolution observations revealed significant sub-structure in most disks that might be hiding dust mass in optically thick regions. The substructure can vary between shallow rings (e.g., TW Hya) to the very clearly separated, narrow rings of disks like AS 209. However, there are also examples of disks without detectable substructure (at current resolution limits) to halt radial drift (e.g., CX Tau \citealp{Facchini2019}). Therefore, the requirement for substructure for the survivability of extended dust emission is not yet clear, requiring even higher resolution observations to confirm.
\end{summarybox}

\subsubsection{\textbf{Gas Surface Density (\Sigg): tracers and limitations}}

Measurements of the gas surface density in disks suffer from similar limitations as the gas mass estimates. Three factors limit the identification of a `global' gas column density tracer: the optical depth of the dust continuum, the optical depth of the gas tracer and the excitation conditions. While the outer disk ($>\!50$~au) is generally optically thin at radio wavelengths, molecular line emission like the CO submm lines are often optically thick for at least part of the disk. The most straightforward solution is then to choose a rare isotopologue line which can be optically thin. The fidelity of this technique for $^{13}$CO ALMA data has been investigated by \citet{Williams2016}, demonstrating its power for large disks that are well resolved (0.2-0.3''). However, \citet{Miotello2018} discuss the difficulties for small disks ($R_c\!<\!100$~au) and also the trade-off between abundant CO isotopologues being optically thick versus the rarer ones showing no emission from the outer disk (where surface densities fall below the hotodissociation threshold/detection limit). Recently, very deep and high spatial resolution ($\sim\,20$~au) observations taken within the MAPS survey \citep{Law2021} reveal that the CO radial intensity profiles ($^{12}$CO, $^{13}$CO, C$^{18}$O, and C$^{17}$O) show less substructure compared to the dust. Combined gas+dust modeling \citep{Zhang2021} is used to derive gas surface density profiles; beyond $\sim\,150$~au, many of these profiles are consistent with power laws and a tapered outer edge, consistent with viscous disk evolution. Another example of a detailed study is the determination of the CO column density in the inner disk (5-21~au) around TW Hya using C$^{18}$O and $^{13}$C$^{18}$O by \citet{Zhang2017}. While the C$^{18}$O line remains optically thick inside of the CO snow line ($<$ tens of au) and thus traces the gas temperature, $^{13}$C$^{18}$O is shown to be optically thin. Using the resolved $^{13}$C$^{18}$O rotational emission, the empirically derived temperatures, and HD $J\!=\!1-0$ to normalize the total gas surface density, they derive a gas surface density of
\begin{equation}
      \Sigma_\mathrm{gas}=13^{+8}_{-5} \times (r/20.5\,\mathrm{au})^{-0.87^{+0.38}_{-0.26}}~\mathrm{g~cm}^{-2}\,\,\,
\end{equation}
within 20~au of the star, demonstrating the power of constraining the distribution of gas with CO even if its absolute normalization factor is difficult to constrain.

The CO submillimeter lines can also become optically thin in the presence of disk cavities, i.e.,\ in transitional disks \citep{Bruderer2013}. After the pioneering work on the Herbig star IRS48 \citep{Bruderer2014}, several works systematically derived the gas column density inside the cavities from $^{12}$CO, $^{13}$CO and C$^{18}$O $J\!=\!3-2$, $6-5$ line data \citep{vanderMarel2015, vanderMarel2016, Dong2017, vanderMarel2018, Gabellini2019}. They find that the drop in gas density inside the cavity is much less than that of the mm-sized dust. The spatial resolution of the data is limited and prevents a firm conclusion on the slope of the gas density profile inside the cavity; however, the line data can be consistent with a positive surface density slope \citep{Dong2017} as expected from models of planet-disk interaction \citep[e.g.][]{Lubow2006}. An additional complication arises from the need to include the effects of gas cooling, which can also diminish the CO emission from inside gaps/cavities \citep{Facchini2018}.

\begin{figure}[th]
      \centering
      \includegraphics[width=\textwidth]{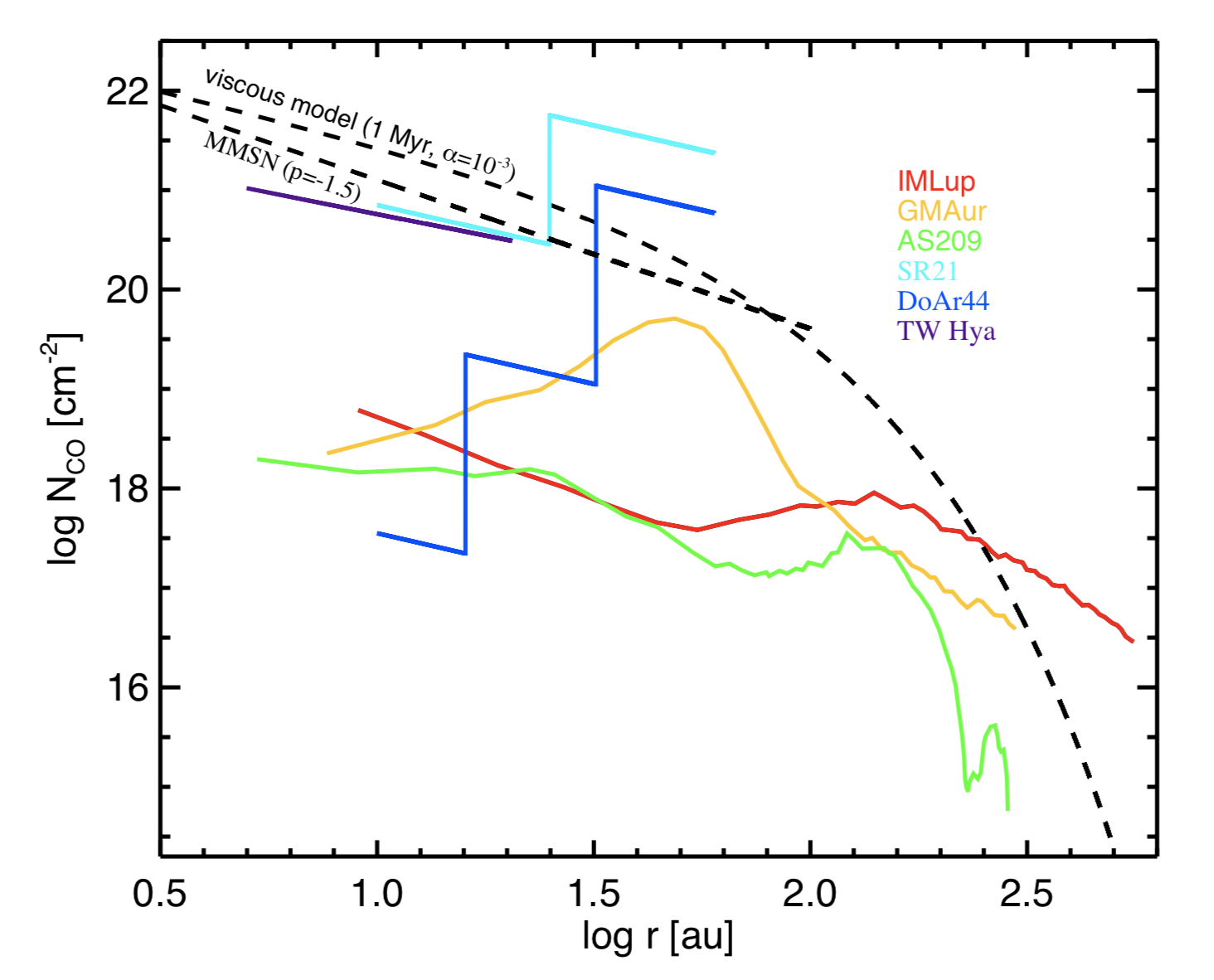}
      \caption{Gas surface density profiles extracted from the literature (CRIRES and ALMA data); the dashed black lines provide the Minimum Mass Solar Nebula (extrapolated to 100~au) and a viscous disk profile for comparison.}
      \label{fig:sigma_gas}
\end{figure}

A complementary approach to determine the gas surface density in the inner disk (inside $\sim\!10$~au) can be taken using the higher excitation CO ro-vibrational lines which trace gas of several 100 up to 1000~K. These lines are regularly detected in disks around T Tauri stars \citep{Najita2003,Salyk2011b}, but to derive reliable surface density profiles, again rare CO isotopologue line profiles need to be detected ($^{13}$CO and C$^{18}$O); this is so far mostly limited to the more massive and bright counterparts, the Herbig disks \citep[e.g.,][]{Carmona2014,Carmona2017}. Figure~\ref{fig:sigma_gas} shows how the warm and cold gas tracers combined can provide a complete picture of the radial distribution of gas inside planet forming disks.

In a few disks with estimates inside 10~au, it has been noted that the gas surface densities/masses derived in the inner disk are too low compared to those derived from the mass accretion rates of these stars \citep[assuming a viscous disk model, e.g.,][]{Woitke2011,Manara2014,Carmona2017}. Solutions to this problem could be that the mass accretion rates are variable and not representative or that the standard viscous accretion model needs to be revised. However, \cite{Costigan2012,Costigan2014} have shown that variability in Class II sources is at most a factor of a few on timescales of months to years; due to asymmetries in the accretion flow.
More secular variability may be in place, but there is no observational evidence for that. Both solutions have been investigated \citep[e.g.,][]{Bai2013, Audard2014} and we refer the reader to the PPVII Chapters by Fisher et al.\ for more details. Of course, another possible interpretation is that depletion of volatile C and O affects the CO emission inside 10~au, which leads us to underestimate the amount of gas that would accrete onto the star. It is therefore important to note that strict measurements or upper limits on the gas surface density in the inner disks are crucial to test our understanding of mass accretion and disk dispersal.

Based on six sources that have been studied in sufficient depth with ALMA (TW\,Hya, IM\,Lup, GM\,Aur, AS\,209, HD\,163296, MWC\,480), several authors found that the gas shows less substructure than the dust \citep{Zhang2017,Zhang2021}; if gas substructure is seen, it does correlate with that of the dust. However, due to the dust opacity potentially being high in the rings, it remains to be quantified how much of the CO substructure is due to varying dust opacity and/or varying gas temperature induced by the dust substructure \citep[e.g.][]{Rab2020}.

\begin{summarybox}
      If it becomes clear that many of the dust gaps seen are not accompanied by equally depleted gas gaps, this would have interesting consequences for understanding the processes that shape the dust, and planet formation scenarios. For example, if the dust rings originate from gas pressure traps and drifting dust grains, the two quantities (brightness of dust continuum emission and gas surface density/pressure) should be related \citep[e.g.,][]{Pinilla2012,Dullemond2018}.
      In the inner disk regions (inside 50~au) overlapping with the Minimum Mass Solar Nebula, the disk observations do not show a simple power law gas surface density profile, but clear substructure \citep{Zhang2021}.
\end{summarybox}

\subsubsection{\textbf{Dust outer radius: measurements and dependence on observed wavelength}}

Defining and measuring the disk outer radius necessarily relies on the disk surface density profile. A physical definition of the disk outer radius \citep[e.g.,][]{Ricci2014,Testi2016} is employed and appropriate in studies of specific sources, but is not the best choice for the characterization of disk sizes in large samples. In such cases the millimeter surface brightness profile, rather than the surface density profile, is usually modeled. Different parametrizations have been adopted in different studies to fit disks brightness profiles: a sharply truncated power law \citep[e.g.,][]{Isella2009, Guilloteau2011}, the self-similar solution
to the viscous evolution equation
\citep[e.g.][]{Andrews2009,Isella2009,Guilloteau2011,Tazzari2017}, and more recently the Nuker profile introduced by \cite{Lauer1995} \citep[e.g.,][]{ Tripathi2017,Andrews2018a,Hendler2020, Sanchis2020}. The Nuker profile is the most flexible approximation, due to the increased number of parameters (five, compared to three). The parameter which controls the sharpness of the transition between inner and outer disc is, however,  hardly constrained in most cases \citep[see e.g.,][]{Tripathi2017}. Therefore a modified version of the self-similar profile, with four parameters, was suggested and successfully used by \cite{Long2019}, \cite{Manara2019}, and \cite{Tazzari2020a}. After modeling the brightness profile, the disk radius is defined empirically as  $R_\mathrm{N}$\%, which is the radius that encircles N\% of the luminosity.
This definition with a fraction of 68\% of the total disk emission has been
used by \cite{Tripathi2017}, \cite{Andrews2018}, \cite{Facchini2019}, \citet{Long2019}, \cite{Manara2019}, and \cite{Tazzari2020a}, while \cite{Tazzari2017} use 95\% of the total disk emission. \cite{Sanchis2020} investigated the quality of $R_{68\%}$ and $R_{95\%}$ as measurements of the characteristic size of the observed disk emission, finding that, given the typical sensitivity of the available observations, the dispersion on $R_{68\%}$is much smaller than on $R_{95\%}$.

Using spatially resolved continuum observations \cite{Tripathi2017}, \cite{Andrews2018}, and \cite{Barenfeld2017} found a tight correlation between the disk sizes and luminosities with a scaling relation $L \propto R^2$ that seems to flatten for older systems \citep{Hendler2020} and to steepen at larger wavelengths, suggesting that large discs are preferentially characterized by a larger spectral index \citep{Tazzari2021}. This relation can be interpreted in different ways: this could reflect the initial conditions at the epoch of disk formation \citep{Isella2009,Andrews2010}, it could be due to the evolution of solids \citep{Tripathi2017,Rosotti2019}, or it may be a consequence of high optical depths \citep{Tripathi2017,Barenfeld2017,Andrews2018,Zhu2019,Long2019,Tazzari2021}. Zormpas et al. (in prep.) have shown, that disks with drift limited dust particles and weak sub-structure naturally follow the size-luminosity relation while disks with stronger sub-structure follow a flatter relation ($F_\mathrm{mm} \propto r_\mathrm{eff}^{5/4}$ instead of $\propto r_\mathrm{eff}^2$, where $r_\mathrm{eff}$ is the effective radius) and that those two relations intersect at the bright end of the observed disk sizes/fluxes. This would suggest that smaller disks on the $r_\mathrm{eff}^2$-size-luminosity relation do not possess substantial (i.e., efficiently trapping) sub-structure, while for bright disks, this distinction cannot be done based only on the disk size and luminosity. Flatter size-luminosity relations were, for example found in the Lupus star forming regions by \citet{Tazzari2021}.

The distribution of smaller dust grains can also be studied by inspecting scattered-light observations \citep[e.g., recent papers of][]{Garufi2018,Avenhaus2018}. In contrast with (sub-)mm continuum studies, the empirical methodology described above has not been employed on scattered-light images. Nevertheless,
from such images it is evident that the micrometer-sized dust grain distribution reaches larger radii than the dust particles responsible for the mm continuum (for example comparing \citealp{Andrews2016} versus \citealp{Rapson2015} and \citealp{vanBoekel2017} for TW Hya or \citealp{Andrews2018} versus \citealp{Avenhaus2018} for IM Lup as shown in \autoref{Fig:IMLup}). One would in fact expect the scattered-light disk radius to be similar to the gas outer radius, as micrometer-sized dust grains are thought to be well coupled to the gas. Quantifying the size difference of micrometer-sized and mm-sized dust grains may be complicated, especially due to the dilution of the stellar radiation field in the outer disk, but future comparisons would be valuable. More details on results from scattered light observations can be found in the PPVII Chapter by Benisty et al..\\

\begin{summarybox}
      In summary, there is probably no correct $R_\mathrm{out}$ definition, but only useful definitions, such as the radius that encloses a chosen percentage of the emission. Among the most popular choices, using the 68\% of the total disk emission is probably the safest approach, as the dispersion on $R_{68\%}$ is much smaller, given the typical sensitivity of the available observations. However, it is important to note that such radius would not necessarily be enclosing the 68\% of the dust mass, as caveats on the the dust optical depth still apply here as in the determination of the dust mass and dust surface density distribution. Measuring disk radii in larger and larger samples of disks, possibly at different evolutionary stages, is becoming a crucial asset to constrain different disk evolution theories. It is therefore important that the radii are measured in a homogeneous manner when performing disk population studies.
\end{summarybox}

\subsubsection{\textbf{Gas outer radius: measurements using different tracers}}

Disk gas size measurements are less common and more difficult, because line emission is faint, especially in the outer regions of the disk. It was only with the advent of ALMA that significant samples of measured gas disk sizes were collected.

Gas disk radii are usually obtained from CO line observations, as they are optically thick for a substantial portion of the disk and are relatively bright also at large separations from the central star. \cite{Ansdell2018} present the first large sample of dust and gas disk radius measurements within a single star-forming region, Lupus. The CO outer radius, $R_\mathrm{CO}$, is measured from the $^{12}$CO moment zero maps \citep[obtained using a Keplerian masking technique to increase the signal-to-noise ratio,][]{Salinas2017} using a curve-of-growth method, in which successively larger photometric apertures are applied until the measured flux is 90\% of the total flux.  This method is applicable only when the CO emission is resolved and the result is that $R_\mathrm{CO} \sim 100-500$ au in Lupus \citep{Ansdell2018}, while some exceptionally large disks that overcome this range. As for the dust component, these are empirical size measurements that are not necessarily directly or simply linked to the gas density distribution. Furthermore, this is only based on the CO emission, that is likely more compact than the actual disk gas outer radius, \Rog, due to ineffective CO self shielding in the outer disk \citep[see e.g., ][]{Miotello2018}. From model predictions, $R_\mathrm{CO}$ should  correlate with \Rog and can therefore be used to constrain disk evolution theories.

\cite{Trapman2020b} have compared measurements of gas outer radii of the brightest Lupus disks from resolved $^{12}$CO emission with viscously expanding thermo-chemical disks models. The observed outer radii can be explained by viscous evolution with $\alpha_\mathrm{visc} = 10^{-4}-10^{-3}$ and small initial radii ($\sim$ 10 au). However, when comparing such model results with CO isotopologue observations, \cite{Trapman2021} find that they do not match the fainter end and non-detections of the observed Lupus population. Viscous evolution seems to be supported also by \cite{Najita2018} who, using a collection of measured gas disk sizes from the literature, showed that older Class II sources have tentatively larger gas radii than the younger Class I sources. Their sample is of course heavily biased towards the brightest objects. Furthermore, their gas disk size measurements were obtained using a variety of different tracers and observational definitions of the gas disk size. All of this warrants a more systematic study. 

The majority of disks targeted by more recent ALMA disk surveys show in fact unresolved or often undetected CO emission \citep{Ansdell2016,Pascucci2016,Eisner2016,Cieza2019,Long2018,Cazzoletti2019,Ansdell2017,Barenfeld2016,Ansdell2020}. Disks with faint CO fluxes may in fact be radially compact. \cite{Barenfeld2017} propose that an explanation for the lack of CO detections in approximately half of the Upper Sco disks with detected continuum emission is that CO is optically thick but has a compact emitting area ($<40$~au). A similar result is found with IRAM Plateau de Bure observations of T Tauri disks by \cite{Pietu2014}, who showed that faint continuum and CO emission in disks often seems to be associated with more compact disks that still have high surface densities in their inner regions. \cite{Pietu2014} also argue that this type of sources could represent up to 25$\%$ of the entire disk population. Furthermore, \cite{Hendler2017} show that the unexpectedly faint [OI]\,63~$\mu$m emission of Very Low Mass Stars (VLMSs) observed with the \textit{Herschel} Space Observatory PACS spectrometer is likely indicative of smaller disk sizes than previously thought. Also, source-specific models based on CO upper limits also lead to the conclusion that some disks need to be compact in size, in order to explain their CO non-detections \citep{Woitke2011,Boneberg2018}. A CN study carried out in the entire Lupus sample has shown that for many of the targeted disks, that also show faint CO fluxes, the critical radius $R_\mathrm{c}$ must be small, even less than 15~au, in order to reproduce the observed low CN fluxes \citep{vanTerwisga2019}. Lupus disks that are not detected in $^{13}$CO emission and with faint or undetected $^{12}$CO emission are consistent with compact disk models \citep[$R_\mathrm{c} \leq 15$~au,][]{Miotello2021}. The fraction of compact disks is potentially between roughly 50\% and 60\% of the entire Lupus sample.

Hence, if a large fraction of the disk population comprised very compact disks, this would challenge viscous evolution theory which would predict extended gaseous disks. Values of the critical radius $R_\mathrm{c} < 15$~au would set strong constraints on the amount of viscosity and cast doubts on whether accretion is driven by viscosity rather than by an alternative mechanism such as magneto-hydrodynamic (MHD) disc winds (See PPVII Chapters by Manara et al.\ and Pascucci et al., for more detail). Other processes should therefore be invoked to truncate disks to such small sizes, as for example external photoevaporation or the encounter with another star (tidal effects). However, we do not expect any of these processes to be relevant in a low mass star-forming region such as Lupus \citep[see e.g.,][]{Winter2018}. An interesting implication for planet formation is that, in such small and optically thick disks, there may be substantial reservoirs of gas for forming Jupiter-like planets within Jupiter's orbital radius. To date, there are no available ALMA observations for a sample of faint Class II disks that are deep enough to discriminate between the two scenarios: radially extended and low mass disks or intrinsically radially compact disks. Deeper observations of $^{12}$CO and $^{13}$CO in fainter disks at a moderate angular resolution are in fact missing and urgently needed.

\begin{summarybox}
      In summary, measuring the outermost extent of gas in disks is more complicated than for the dust. This is due to various reasons: deep enough gas observations are more expensive and therefore rarer; furthermore the outer gaseous extent depends on the choice of the molecular tracer (traditionally $^{12}$CO), whose connection with the actual gas outer radius is not trivial. It would be critical to build a homogeneous sample of measured disk gas outer radii for a large fraction of the known disks, as the small number statistics is currently the most limiting factor in studies of disk evolution.
\end{summarybox}

\subsubsection{\textbf{Gas versus dust outer radii}}

Gas disk outer radii, as probed by the bright $^{12}$CO emission, are generally larger than dust radii, probed by sub-mm continuum emission. This difference could be due to optical depth effects, with $^{12}$CO lines being more optically thick than the continuum emission \citep[see e.g.,][]{Dutrey1998,Guilloteau1998,Panic2009}. Fitting both dust and gas tracers with the same surface density profiles \citep{Hughes2008,Andrews2009} is appropriate for low-resolution, low-sensitivity observations, but it does not work for
more recent observations with higher sensitivity and angular resolution. It was in fact revealed with ALMA, that in some sources the dust outer edge decreases too sharply with radius, and cannot be reproduced by a tapered outer disk \citep[e.g.,][]{Andrews2012,Andrews2016, deGregorio-Monsalvo2013,Cleeves2016}.

Grain growth and consequent radial drift of large particles from the outer to the inner disk could be an explanation for the observed behavior. Multi-wavelengths continuum observations suggest that dust particles can grow to at least mm/cm sizes in protoplanetary disks \citep[see e.g.,][]{Testi2003,Natta2004,Lommen2007,Lommen2010,Andrews2005,Ricci2010a,Ricci2010b,Testi2014,Andrews2015,Birnstiel2016}. Furthermore, there is increasing evidence of a decreasing gradient of dust size as a function of distance from the star, as expected from models \citep[see e.g.,][]{Miotello2012,Guilloteau2011,Perez2012,Perez2015,Menu2014,Tazzari2016}. Further theoretical modelling supports this idea \citep[see e.g.,][]{Birnstiel2010,Birnstiel2012}; in particular, \cite{Birnstiel2014} have shown that the  observed sharp edge is also predicted by dust evolution models. This outer edge of the disk might however occur at such low dust surface densities, that the emission is below the current sensitivity. Alternatively, the outer edge of the dust could be, in some cases, a result of an opacity feature (see \autoref{sec:gas_mass}, which would also explain the size luminosity relation \citep{Rosotti2019}).

To answer the question about the importance of radial drift, \cite{Facchini2017} have coupled dust evolution models to physical-chemical models of protoplanetary disks, focusing on the radial properties of continuum and CO lines. Such models show that differences in gas and dust radial extents can be largely explained by the difference in optical depth of gas versus dust, without the need to invoke radial drift. The latter, on the contrary, primarily affects the shape of the outer edge of the (sub-)mm continuum intensity profile, which is steeply truncated, in particular for low values of turbulence. However, using similar models, \cite{Trapman2019} more recently showed that if $R_\mathrm{CO}$ is larger than 4 times $R_\mathrm{dust}$, this morphology suggests  that radial drift being the main process setting the gas-to-dust disk size ratio.
Recently \cite{Toci2021} have modeled the disk evolution, considering viscosity, grain growth, and radial drift. They have found that, $R_\mathrm{CO}/R_\mathrm{mm-dust}$ becomes large by a factor of more than 5 after only 1 Myr. In their models, this is due to radial drift, but it is inconsistent with available measurements in nearby SFRs, where the ratio is much smaller. The authors conclude that substructures, commonly invoked to stop radial drift in large, bright disks, must then be present but unresolved, in most disks.

Finally, stellar multiplicity plays an important role on the dust and gas disk extent. In fact, the gas outer radii of disks surrounding stars in multiple systems are expected to be truncated to sizes that are a fraction of the distance between the two objects, with a dependence on the orbital eccentricity, the stellar mass ratio, the viscosity and temperature of the disks, and their co-planarity \citep[e.g.,][]{PapaloizouPringle1977}.
      Surveys of multiple stellar systems conducted with SMA and ALMA show that discs in multiple systems are on average fainter at any given stellar mass than those in single systems \citep{Harris2012,AkesonJensen2014,Akeson2019,Manara2019,Long2019}. Recently \cite{Rota2022} presented deep ALMA observations of CO line and continuum emission in ten multiple stellar systems in the Taurus SFR. They compared the  CO radii enclosing 68\% and 95\% of the total disc fluxes and compared to the dust radii estimated by \cite{Manara2019}, and
      derived the gas-to-dust size ratio to be $\sim 2 - 4$. Unexpectedly, the effective (68\%) gas-to-dust size ratio is  statistically compatible to the ratio estimated by \cite{Sanchis2021} for a population of single disks. When considering the 95\% disk radius, instead, the gas-to-dust size ratio is found to be on the high-end of the values found for isolated systems, possibly due to the sharp truncation of the outer dust disk expected in binary systems. Therefore, the difference between the gas-to-dust size ratio in single and multiple disks does not seem to be so clear.

\begin{summarybox}
      In conclusion, our knowledge of the disk gas outer radii, and accordingly of the \Rog/\Rod, are based on $^{12}$CO line and sub-mm continuum emission. Different types of modeling efforts, including chemistry and/or dust evolution and grain growth, have been made, but not all studies come to the same conclusion concerning viscous evolution of gas in the presence of radial drift of dust. It is anyway clear that dust substructures have a clear role in this discussion.
\end{summarybox}

\subsubsection{\textbf{$\Td(r)$ from multi-band dust observations}}

Determining the disk thermal structure is fundamental, as it controls the chemical composition of the disk (e.g., ice lines, endothermic reactions), dust dynamics (radial migration and settling) and gas instabilities. More directly, it sets the molecular excitation conditions, and thus the emission line intensities, and the  dust continuum fluxes.

According to theoretical models, the dust temperature distribution shows an increasing vertical gradient and decreases radially outward \citep[see top-left panel of Fig. \ref{fig:pp7disk},][]{Calvet1991,Kenyon1987}. This is the result of the stellar irradiation as well as the vertical distribution of the dust grains. The dust temperature is primarily set by the stellar irradiation intercepted by the grains. The star light is absorbed or scattered by small grains present in the upper layers of the disk, which then re-radiate towards the mid-plane \citep[see e.g.,][]{Chiang1997,D'Alessio1998,Dullemond2007}. In the context of planet formation it is particularly interesting to constrain the dust temperature in the disk mid-plane, where the bulk of the gas and dust mass are, which then determine the formation mechanism and composition of planetary cores.

The classical approach to determine the dust temperature is to forward model the infrared Spectral Energy Distribution (SED). After assuming a disk surface density profile and the dust opacity, such models simulate the the energy propagation throughout the disk and generate synthetic SEDs to be compared with the observations. This process is iterated until the observed SED is reproduced. This approach is quite effective, but prone to large degeneracies \citep[][]{Thamm1994, Heese2017}. Combining SEDs with spatially resolved observations can help reduce the uncertainty, as done for example by \cite{Pinte2008}. Especially combining continuum images in multiple bands, optically thick and thin emission, can allow the combined determination of dust temperature and dust properties \citep{Kim2019}.

\begin{figure}[ht]
      \includegraphics[width=\hsize]{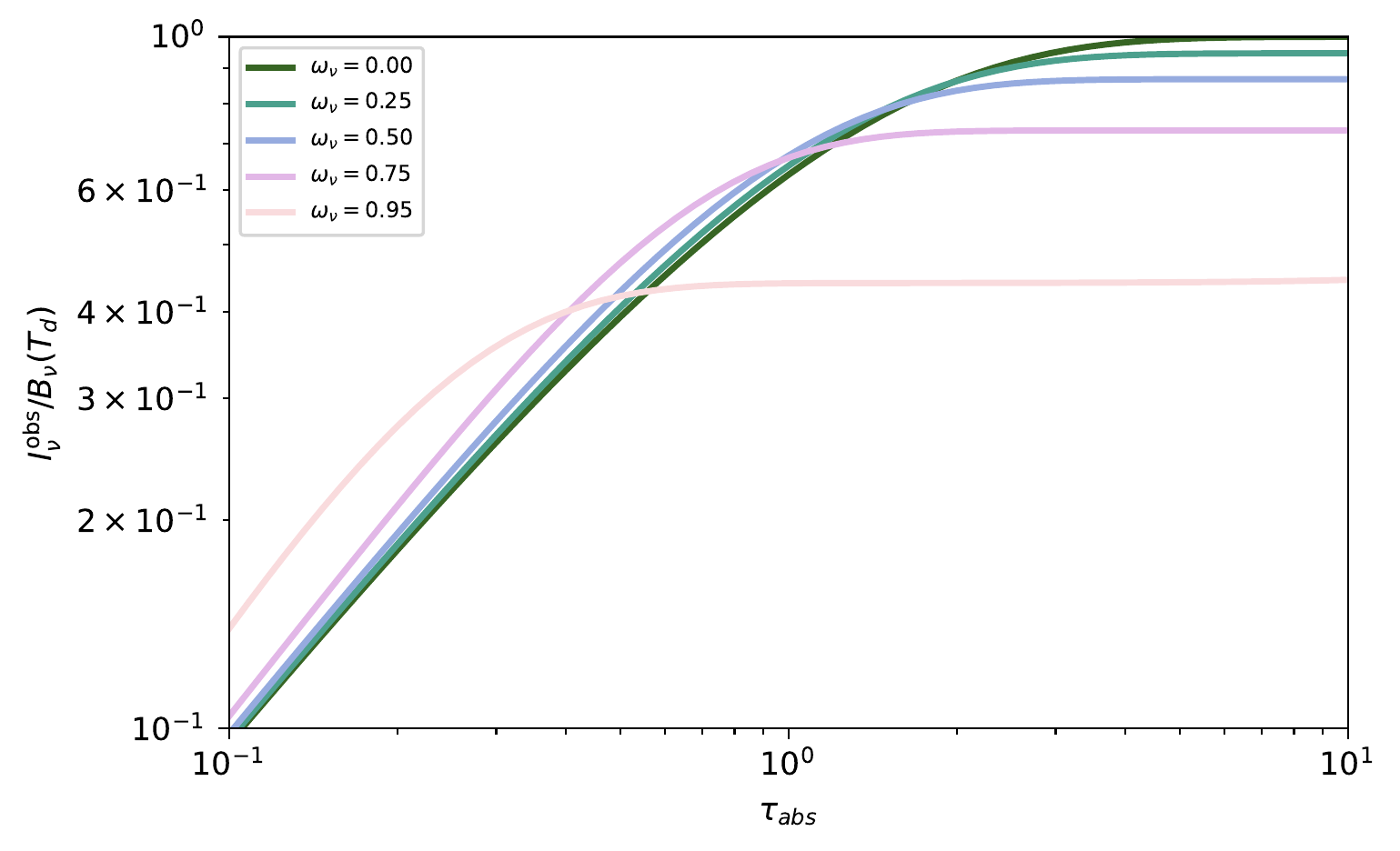}\vspace{-3mm}
      \caption{Emitted intensity from a homogeneous slab model as function of absorption optical depth for different values of albedo $\omega$ (after \citealp{Miyake1993,Zhu2019}).}
      \label{fig:scattering}
\end{figure}

Spatially resolved images allow measuring the brightness temperature of the disk which, in the optically thick case (cf. \autoref{eq:thin_emission}) and for low albedo is identical to the dust temperature in the disk photosphere. The effect of scattering on the observed continuum intensity is shown in \autoref{fig:scattering} as a function of the \textit{absorption} optical depth. For increasing values of the albedo, the extinction (scattering + absorption) optical depth becomes larger \citep{Miyake1993, Zhu2019}. It can be seen that for optical depth between 0.1 and 1, scattering can slightly increase the emission as the longer effective path allows the radiation to better thermalize. For optical depth above unity, the effect of scattering is to reduce the emitted intensity. This makes the disk appear colder if scattering is not included. However, most continuum radiative transfer codes routinely include scattering when computing images and SEDs \citep[see e.g.,][]{Pinte2009}; the effect of scattering has been often neglected for simplicity in the direct interpretation of sub-mm continuum observations such as deriving spectral indices \citep[see e.g.,][]{Williams2011}. Using both line and continuum observations can help to measure the disk temperatures as discussed in Sect.~\ref{sec:Tgas-radial-profile}. Alternatively, if the temperature is known, combined gas and dust observations can be used to constrain the albedo of the dust particles \citep{Guilloteau2016,Isella2018}.

Combining interferometric data from near- to mid-IR, it will become possible to constrain the dust temperature gradients in the surface of the inner disk also for T Tauri disks; this has been done recently for the Herbig disk HD\,179218 (PIONIER, MATISSE, MIDI data) showing the need for a radial dust temperature gradient inside 10~au \citep{Kokoulina2021}.

\begin{summarybox}Interestingly enough, it seems that the dust temperature structure has been studied in much less detail, compared with the gas temperature (see Sec. \ref{sec:Tgas-radial-profile}). However, this does not imply that $T_\mathrm{dust}$ is less of an interesting disk property, as in fact it affects for example the importance of disk instabilities. Therefore, there is still much theoretical and observational work to be done on dust temperature structure.
\end{summarybox}

\subsubsection{\textbf{$\Tg(r)$: constraints from line excitation and chemistry}}
\label{sec:Tgas-radial-profile}

There are direct methods to determine the gas temperature such as gas emission line fluxes, radial emission profiles, and channel maps and also indirect (chemical) methods such as molecular emission maps tracing specific snow lines. In the latter case, the gas emission always traces specific radial and vertical regions of the disk, thus a single tracer allows one to reconstruct the gas temperature profile in only a specific part of the disk (namely where it predominantly emits from). Multiple tracers are needed to construct a complete picture, ideally with overlapping coverage. With this in mind, we discuss the direct gas line methods first grouped by radial origin from the inner disk (near- to mid-IR, \SIrange{1}{10}{au}), to the outer disk (submm, beyond \SI{10}{au}). The far-IR observations fall in between covering gas lines of both high and low excitation.

\textbf{Inner disk (\SIrange{1}{3}{au}) --- terrestrial planets:} The first estimates of typical inner disk gas temperatures originated from CO, OH and water emission lines in the near-IR \citep[see PPV review of][]{Najita2007} and were in line with our theoretical understanding from thermo-chemical disk models. Additional confirmation came from the slab analysis of Spitzer spectra \citep{Carr2008, Pontoppidan2010a, Salyk2011a}, again confirming that typical gas temperatures in the disk surface at distance up to a few au are \SIrange{900}{500}{K}. There is quite some range in \Tg from different mid-IR molecules, e.g.\ water is best fit with temperatures around 500~K, while CO and OH show a large range of temperatures across different disks ($500-1700$~K). CO$_2$ and HCN reside in between with typical temperatures around $700-800$~K. Since LTE slab models have a number of free parameters (gas temperature $T$, column density $N$, emitting area $A$), these results are somewhat degenerate and suffer from assumptions like co-spatiality in the absence of spatially resolved spectral line profiles. From a thermo-chemical modeling perspective, we expect the different molecules to originate from different heights in the disk \citep{Woitke2018}, ordered by height: OH, CO, H$_2$O, CO$_2$, HCN (from high to low height); thus finding different temperatures for those molecules is consistent with detailed disk models. In general, mid-IR observations probe only the column of gas above the local dust continuum. In addition, molecules such as water can be efficiently pumped by thermal dust emission in the inner disk \citep{Kamp2013} and thus the measured water temperature could rather reflect the typical dust temperature at a few au.

With the help of spectral line profiles, the emitting area is better constrained and especially CO ro-vibrational lines have been used to derive gas temperature profiles inside few astronomical units  \citep{Salyk2011b, Bast2011, Brown2013}. Rotational diagrams result in temperature estimates of $700-1700$~K for typical emitting regions of $\leq\!0.2$~au. While UV fluorescence can affect the vibrational temperature close to the star \citep{Brittain2007,Brown2013}, the rotational temperature should be in LTE and so reflect the gas temperature. \citet{Bast2011} focus on broad single-peaked CO line profiles and find indeed that the vibrational temperatures are higher ($\sim\!1700$~K) than the rotational temperatures ($300-800$~K), suggesting that fluorescence likely affects the vibrational excitation. Interestingly, the $^{13}$CO rotational temperature is systematically lower than $^{12}$CO and the line profiles are narrower for sources where those lines are detected \citep{Brown2013}. This is consistent with an origin at larger radial distances. For much smaller samples, spectrally resolved profiles of water and HCN have been observed \citep{Pontoppidan2010b, Mandell2012, Najita2018}. Those line profiles are again consistent with the emission originating inside $\sim\!2$~au and the line profiles of water, HCN and C$_2$H$_2$ in the 12.4 and 12.8~$\mu$m wavelength region match; contrary to non-LTE disk model predictions \citep[IR pumping][]{Bruderer2015}, the line profiles of HCN at 3 and 12.4~$\mu$m match. However, water lines of different excitation cannot be matched within a single LTE slab model.

With the new VLT instruments GRAVITY and MATISSE and the upcoming ELT/METIS instrument, the study of spatially resolved gas line emission becomes feasible also for T Tauri disks. Previous work has focussed on the atomic gas (Br$\gamma$ line) on sub-au scales in Herbig disks with VLTI/AMBER \citep[e.g.,][]{CarattioGaratti2015, GarciaLopez2015}. This can break degeneracies and settled discussions on the inner disk gas structures.

To conclude, the estimate of accurate disk temperatures from near- to mid-IR lines is limited due to the degeneracies in molecule location and excitation processes (e.g.,\ the role of IR and UV pumping). However, typical temperatures inferred in the inner disk are largely compatible with the resutls of thermo-chemical disk models.

\textbf{Middle disk ($3-10$~au) --- gas giant planets:} Gas temperatures in this region are estimated from very different wavelength ranges. Some of the low excitation water lines in the mid-IR likely probe this region (e.g., water at $33~\mu$m), but also the higher $J$ rotational lines of CO probed in the far-IR ($J_\mathrm{up}\!>\!9$). Using a power law disk model, \citet{Fedele2016} fit the full CO ladder for TW\,Hya and found a steep temperature decline $r^{-0.7 \ldots 0.8}$. It is important to note that while such a power law behavior is expected in the midplane of the outer disk, the surface layers and inner disk can deviate from such a simple profile due to the gas/dust temperature decoupling and radiative transfer effects \citep{Kamp2004, Pinte2009, Fedele2013}. \citet{Zhang2013} used an indirect method to derive a temperature anchor in the disk around TW\,Hya, the water snow line location. From an emission line ``drop-out'' method using mid-IR to far-IR water lines, they deduced the surface snow line location to be at 4.2~au \footnote{However, it should be noted that using snow line locations to pinpoint temperatures will inherently be subject to changes in accretion rate, either over the disk secular evolution or as a result of bursts. This will most strongly impact close in snow lines, like that of water.}. 

Even ALMA is pushed to its limits when trying to spatially resolve the gas emission inside 10~au at typical distances of star forming regions. The \emph{Herschel} satellite lacked the sensitivity to detect routinely CO, H$_2$O and OH cooling lines from class II disks \citep{AlonsoMartinez2017}. This and the lack of very sensitive far-IR satellites (with mission concepts like JAXA and ESA's SPICA having been cancelled) strongly limits our knowledge about the thermal structure of the region where gas giant planets similar to Jupiter may be forming.

\textbf{Outer disk ($10-$few\,100~au) --- the outer planets and beyond:} The far-IR wavelength range is rich in gas cooling lines of abundant species (e.g.,\ OH, H$_2$O, CO). \citet{Kamp2013} find that the far-IR lines from the disk around TW\,Hya are very sensitive to temperature (including also [O\,{\sc i}] and water). The UV excess and X-ray emission from T Tauri stars (in conjunction with disk flaring) is key in explaining the brightness of the [O\,{\sc i}]\,63~$\mu$m line \citep{Pinte2010}, and thus the temperature of the disk surface at distances of $30-100$~au. Any kind of shadowing by an inner disk (e.g., puffed up geometry, misaligned inner disk) will affect the temperature in the outer disk surface and thus the line emission from there \citep{Woitke2019}.

Spatially resolved channel maps of molecular lines provided by interferometers at submillimeter wavelengths can be used to infer radial gas temperature profiles in the outer disk either directly or through modeling. For example, using power law disk models, \citet{Chapillon2012} used CN and HCN observations with the Plateau de Bure Interferometer (PdBI) to fit -- among other parameters -- the gas temperature profile. However, CN and HCN do not give consistent profiles and also the two sources differ significantly (from flat, $q\!=\!0$ to steep $q\!=\!0.95$), suggestive of different radial or vertical distributions of these molecules. More recently, combining 2D thermo-chemical disk modeling with ALMA observations of CN, HCN and HNC have been employed to understand the how disk irradiation (UV excess, geometry) and gas temperature relate to break the remaining degeneracies \citep{Cazzoletti2018, vanTerwisga2019, Long2021}. In the case of the face-on disk TW\,Hya, a series of studies have aimed to disentangle the disk temperature and surface density profiles using a series of CO isotopologue lines \citep{Schwarz2016, Zhang2017, Huang2018a}. These studies have relied on the high optical depth of CO and its isotopologues in the inner disk to derive the brightness temperature. In the disk around TW\,Hya, \citet{Schwarz2016} derive from $^{13}$CO and C$^{18}$O data a steep temperature profile inside 30~au followed by a flat plateau ($\sim\!20$~K) out to $\sim\!65$~au. The latter suggests that the authors detected the CO iceline surface. \citet{Zhang2017} used the spatial distribution of the even rarer isotopologue $^{13}$C$^{18}$O to estimate the midplane CO iceline location to be $20.5\pm1.3$~au. The midplane gas temperature inside that CO iceline has a power law index of $\sim\!-0.5$. Alternatively, for high spatial resolution data, peak brightness maps of a single optically thick CO line can be used to derive radial disk temperature profiles \citep{Weaver2018}.

\citet{Qi2013} used the fact that N$_2$H$^+$ is destroyed effectively in the presence of CO to deduce the location of CO freeze out. Using the size of the emission ring observed in  TW\,Hya, they inferred the CO iceline to be at $\sim\!30$~au. Given the measured binding energy of CO on CO \citep{oberg05}, this phase transition would translate into a calibration anchor point of 17~K. However, some uncertainty exists in the interpretation of the emitting layer of N$_2$H$^+$. \citet{Nomura2016} derived the gas temperature in the disk around TW\,Hya from ALMA observations of various molecules. They show that N$_2$H$^+$ has a higher temperature than $^{13}$CO and the dust. This measurement either suggests N$_2$H$^+$ emits from closer to the disk surface and is thus not a good midplane temperature tracer or that the $^{13}$CO line is optically thin and the brightness temperature does not reflect the gas temperature. In another study of TW Hya using multiple N$_2$H$^+$ transitions, \citet{Schwarz2019} confirmed a warm excitation temperature higher than the CO freeze out temperature. Subsequent disk modeling by \citet{vantHoff2017} confirmed the necessity of forward modeling approaches to derive the CO iceline location from N$_2$H$^+$ observations. It is clear from these observational and theoretical studies that the radial temperature profile is difficult to disentangle observationally from the vertical temperature gradient, and both must be simultaneously constrained, as further discussed in Section~\ref{sec:vertT}.

\begin{summarybox}
      Very little work has been done on the quantitative interpretation of temperatures derived from gas observations. This is partly due to the lack of high quality gas data and the need for coherent multi molecule/line datasets. This gap has started to be filled with data from the recent MAPS ALMA Large program \citep{oberg21} and this effort will likely continue with future observations.
\end{summarybox}

\subsection{\textbf{Vertical structure}}\label{sec:vertstruct}

Protoplanetary disks have finite thickness due to the pressure support from the gas disk. The vertical structure is sensitive to a number of factors, specifically the gravitational force exerted by the central star, the dominant heating and cooling agents, and the degree of dust evolution (which impacts the thermal structure). In addition, dynamical factors may play some support, either through strong mixing or support from magnetized winds (see PPVII Chapter by Pascucci et al.). In this section, we summarize our current knowledge of the vertical distribution of both solids and gas and how these contribute to our knowledge of planet formation.

\subsubsection{\textbf{Observational constraints on dust settling and dust scattering surface}}

\label{sec:obs_settling}
The vertical distribution of dust is determined by a balance of multiple factors. The gas and any associated turbulent stirring motions lift the dust upward, while gravity tends to sediment grains toward the midplane. In this section we review our current understanding of observational constraints on the vertical distribution of dust grains.

The vertical distribution of dust grains is intimately linked to the properties of the gas. As discussed, the vertical structure of the gas is basically understood by the force balance between vertical pressure gradient and the gravitational force toward the midplane, which gives us the gas scale height, $h_\mathrm{g} = c_\mathrm{s} / \sqrt{G\,M_\star/r^3} = c_\mathrm{s}/\Omega$.  The vertical distribution of the dust strongly depends on how well coupled grains are to the gas: dust grains that are well-coupled to the gas should follow the gas profile, while dust grains decoupled from the gas tend to settle toward the mid-plane \citep[e.g.,][]{Dubrulle1995}.

\begin{figure}
      \includegraphics[width=1\textwidth]{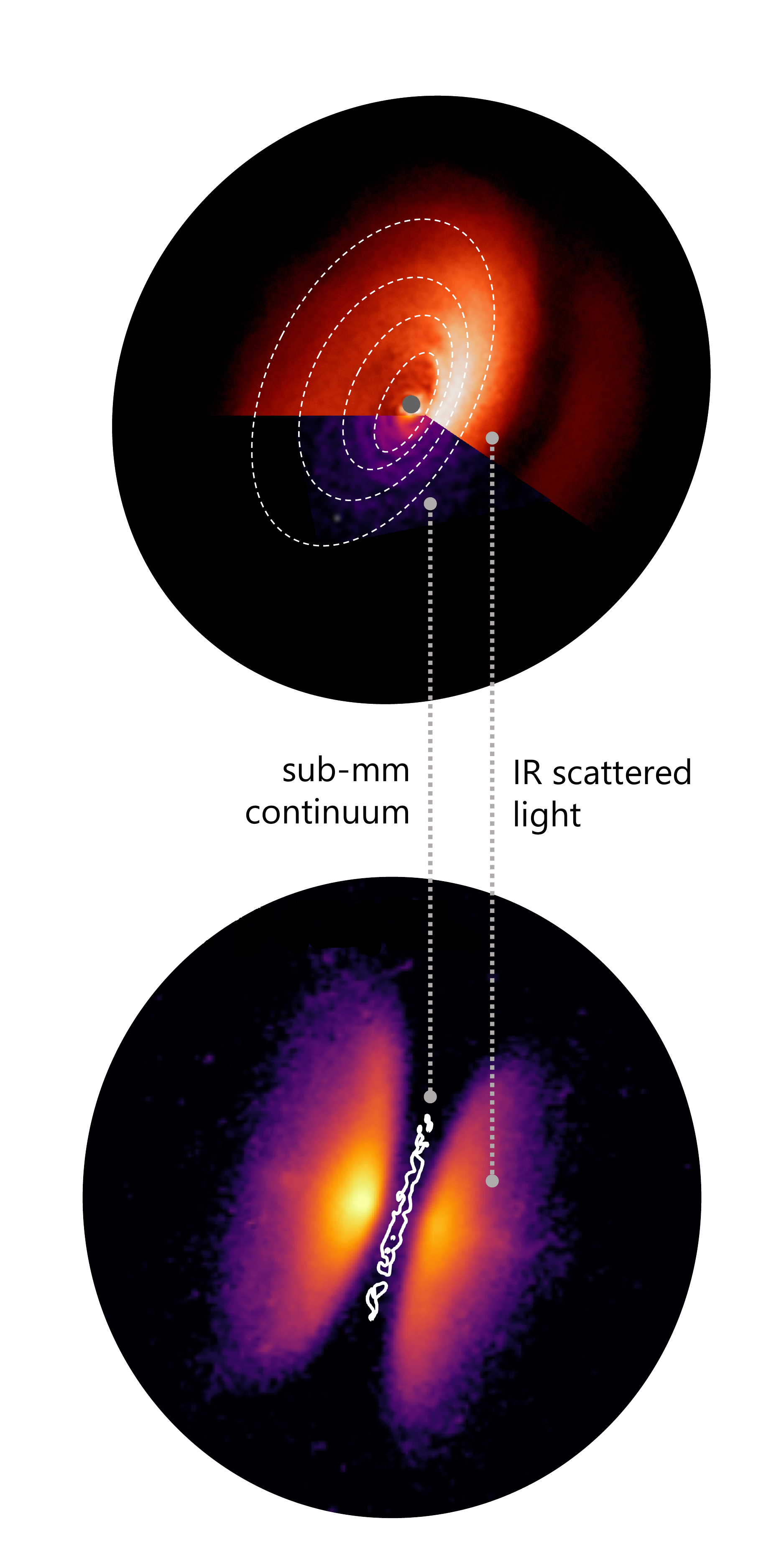}\vspace{-5mm}
      \caption{\emph{Top - }Collage of observations of the disk around IM Lup in different tracers: scattered light from small dust on the top \citep{Avenhaus2018}, and sub-mm continuum from mm-sized grains in the bottom \citep{Andrews2018}. The white dashed lines guide the eye to the disk surface layer, seen in scattered light, while the continuum emission at mm wavelengths shows a flat and settled midplane. \emph{Bottom - } Combined observations of the edge-on disk Tau 042021 \citep{Villenave2020}. The scattered light emission (colored map) is flared and vertically extended, while mm emission (white contours) trace an extremely flat midplane.}
      \label{Fig:IMLup}
\end{figure}

The coupling of dust particles depends on the size of the grains as encapsulated by their Stokes number (which is approximately the ratio of the time scale required to adjust to the gas velocity, and the orbital time scale). Variations in the vertical distribution with grain-size is reflected in the appearance of infrared and millimeter-wave observations, where the optically thick infrared observations trace stirred-up micron-sized dust grains, while mostly optically thick millimeter-wave observations trace settled millimeter-sized dust grains close to the mid-plane.
This is visually demonstrated by comparing the images taken by ALMA SPHERE, and HST \citep[see IM Lup and Tau 042021 in \autoref{Fig:IMLup};][]{Avenhaus2018,Andrews2018,Villenave2020}.

The infrared scattered surface has been imaged at several individual targets with facilities such as HiCIAO \citep[e.g.,][]{Akiyama2016}, GPI \citep[e.g.,][]{Rapson2015,Wolff2016}, and SPHERE \citep{vanBoekel2017,Garufi2016, Avenhaus2014a, Avenhaus2014b, Stolker2016, Pohl2017}
A systematic survey of disks' infrared surface layer has been performed with SPHERE \citep{Avenhaus2018}, which shows that the aspect ratio is an increasing function of the distance from the central star. They found that the scattering surface for 5 disks is well represented by $z/h = 0.16\, (r/\SI{100}{au})^{1.22}$. This is in broad agreement with theoretical predictions of a flaring disk model \citep[e.g.,][]{Kenyon1987,Chiang1997}.

The millimeter-wave continuum emission is well modeled with settled dust grains even in the pre-ALMA era \citep[e.g.,][]{Grafe2013}.
With ALMA, there are several lines of evidence of dust settling in millimeter-wave images.
An edge-on disk survey with ALMA enables us to directly compare the vertical structures, showing that the optical-NIR vertical scale is larger than the millimeter-observed disks, which is the direct evidence of dust settling \citep[see Fig. \ref{Fig:IMLup},][]{Villenave2020}.
Substructures also hint the dust settling status.
High spatial resolution images of inclined axisymmetric rings enable a geometrical estimate of the vertical extent of the rings.
HL Tau is consistent with a geometrically flat disk \citep{Pinte2016}, which indicates strong dust settling.
However, this is not always the case; HD 163296 has both puffed up and settled rings \citep{Doi2021}.
This is also tested by comparing the dust ring width and the gas scale height, which is also determined by the balance between gas diffusion and dust decoupling \citep{Dullemond2018, Rosotti2020}.

In this way, a consistent picture of the dust vertical structure is emerging -- the infrared emitting surface is puffed up while millimeter-wave emitting regions are concentrated close to the mid-plane.
However, there have been some questions proposed.
Some disks show near infrared scattered brightness is lower than expected \citep{Mulders2013, Muro-Arena2018}.
HD 100546 shows low albedo, which can be reproduced by assuming strong forward scattering \citep{Mulders2013}.
HD 163296 shows a deficit of infrared scattered light at the outer disk, which requires small dust grains depleted at the outer disk.
There are several ways to make the infrared scattered light appear fainter.
One possible way is to assume the smallest grains to be as large as a micrometer.
If the dust grains follow a typical power-law size distributions, where $n \propto a^{-3.5}$, the smallest dust grains dominate the infrared opacity.
Thus, if the smallest grains have the size of a few microns, which has low albedo at the infrared wavelengths.
Another way is to include the porosity or shape of dust grains \citep{Kirchschlager2014, Min2016, Tazaki2019}.
However, this raises another question: why are the micron sized grains puffed up?
Molecular line emissions have shown that disk turbulence (as measured by the $\alpha$ parameter, such that typically $\sqrt{\alpha}\,c_\mathrm{s}$ is taken to be the turbulent velocity dispersion of the gas) is very weak in the outer disk \cite[e.g., in TW Hya $\alpha < 0.007$, in HL Tau $\alpha=3\times10^{-4}$, see][]{Teague2016, Flaherty2018,Pinte2016} except for DM Tau. In the inner disk (within 10 au) turbulence can be measured using near-IR molecular lines, but this was so far possible only for one Herbig disk, that showed a much higher alpha in the inner disk, compared to the typical measurements found from ALMA for the outer disk. In this matter, the role of the Extremely Large Telescope (ELT) will be to provide higher spatial and spectral resolution than \emph{James Webb Space telescope} (JWST) such that at $<10$ AU scales are reachable at high spectral resolution for more than just the most extreme disks accessible with existing telescopes \citep{Jang-Condell2019}.

Such small values of the turbulence parameter, as measured by ALMA, may not be sufficient to explain the vertical distribution of micron sized dust grains. Detailed modeling of the vertical structure (and the turbulent properties at different heights above the mid-plane) will be needed for a future understanding of how dust grains grow and are vertically distributed at different locations in the disk and might need more complicated ingredients. This lack of turbulence would not be able to explain observed accretion rates and might indicate that some other mechanism is driving the evolution of the disk. This would require a mechanism that is transporting mass and angular momentum without causing gas turbulence above the observable levels and without mixing the dust mid-plane significantly. This could happen through magnetic winds or current sheets (see for example \citealt{Bai2017}) or dust stirring by planets through meridional flows \citep{Binkert2021,Bi2021}.

\begin{summarybox}
      In summary, combining IR scattered light and sub-mm continuum observations it is generally found that smaller grains follow the gas vertical distribution to very elevated regions of the disk, while mm-sized grains are recluse to thin and settled midplanes. This would imply that disks are highly turbulent, while line emission studies with ALMA have found the opposite, with strong limits on non-Keplerian, non-thermal motions. As of today, this is still left as an open issue.
\end{summarybox}

\subsubsection{\textbf{Vertical temperature stratification and constraints on geometry of line emission}}\label{sec:vertT}

As discussed in Section~\ref{sec:Tgas-radial-profile}, gradients in disk temperature are expected both in the radial and vertical directions. Observational evidence for the presence of vertical temperature gradients has become much stronger in recent years; however, measuring and interpreting robust numbers from observations is challenging. This is due to the fact that vertical temperature gradients are intractably linked to many other fundamental properties, such as the dust evolutionary state (see Section~\ref{sec:dust_properties}) and the disk chemical composition \citep[e.g., through heating and cooling agents;][]{kamp01,gorti04,woitke09,Bruderer2012,grassi20}. In addition, disk inclination and related projection effects muddy direct observational temperature constraints, often requiring either very high spatial resolution data \citep{Rosenfeld2013,Pinte2018,Law21_MAPS_surfaces_vertical_distributions} or forward modeling approaches \citep[e.g.,][]{Fedele2016,Zhang2017}. These factors, among others,  make vertical temperature gradients intrinsically difficult to observationally constrain, however as discussed in this section, much progress has been made since PPVI.

Across the literature, many different models, both analytic and Monte Carlo in nature, have been constructed to predict temperature profiles of irradiated accretion disks
\citep[see e.g.,][]{Bruderer2012,Woitke2019}. Models have varying levels of sophistication, including layered dust settling, accretion heating, thermochemistry, and feedback with the vertical gas density profile. Specifically, the gas scale height is linked to the temperature through gas pressure and magnetic support \citep[e.g.,][]{bai2009,hirose2011} against gravitational forces from the star and disk. For example, the enhanced pressure support in warmer disks results in larger scale heights than for cooler disks. Furthermore, a more puffed up vertical structure intercepts more heating radiation from the star, enhancing the pressure support balanced by increased surface area for cooling, resulting in a locally elevated vertical structure. Disk substructure, such as inner holes or radial gaps, can change how the disk is illuminated, further changing the vertical profile of the disk locally, e.g., creating puffed up inner rims \citep[e.g.,][]{Natta2001,Isella2005,Binkert2021,Bi2021} or outer edges of disk gaps \citep[e.g.,][]{jangcondell12}. These effects are important for understanding the degree of deviation from a standard simplifying assumption of a disk being locally vertically isothermal, a common assumption resulting in a Gaussian vertical density profile. This assumption is however often still adopted in models of disk structure, and within gaps as discussed in \cite{vanderMarel2018}, because of its simplicity.

Early observational constraints on vertical gradients in disk temperature have primarily focused on thermal continuum and optically thick gas. Detailed fitting of the SED demonstrated that a warm irradiated surface was needed to reproduce the IR excess and the 10 micron silicate feature, however, the details of the temperature were sometimes difficult to disentangle from other disk properties such as composition and/or structure \citep[e.g.,][]{Woitke2011,Woitke2019}. Optically thick CO emission has a been a widely adopted temperature tracer, both through CO ladder studies, e.g., high $J$ CO \citep[e.g.,][]{Fedele2016} and inferred from T$_\mathrm{gas}$ derived from multi-isotopologue studies ($^{12}$CO vs. $^{13}$CO), where the latter is expected to trace deeper layers in the disk due to its lower abundance and corresponding optical depth \citep{Pietu2014,Dutrey2017,Zhang2017,Pinte2018,Schwarz2019}.

Observational temperature constraints in disks are becoming increasingly more sophisticated with the advent of high resolution observations of gas and dust afforded by interferometers. Interferometers have provided the necessary fine-grained resolution to not only map radial temperature gradients, but to also {\em split} the layers of the disk to determine temperatures more locally as was first done in \citet{Rosenfeld2013} for the HD 163296 disk  \citep[see also][\emph{Paneque-Carre{\~n}o et al.}, subm.]{Pinte2018}. In the following paragraphs, we outline our modern observational understanding of the temperature of the elevated molecular layer and that of the midplane.\\

\textbf{The molecular layer} of the disk -- the region between which molecules are shielded from dissociating radiation but are warm enough to remain in the gas phase -- has the most extensive amount of data on its temperature in particular for the outer disk at $T < 100$~K. This abundance of data is due to the similar temperature of the layer to the energies required to excite rotational emission for a number of common interstellar molecules with transitions at submillimeter wavelengths. The temperature of this layer can be determined through a number of different, complimentary techniques, invoking (1) chemistry, (2) line excitation, and (3) line optical depth.

The first of these techniques, chemistry, utilizes the simple fact that molecules can freeze out onto dust grains at distinct dust temperatures \citep[e.g.,][]{collings2004}. Each molecule freezes at a particular temperature; however, the specific temperature can vary slightly depending on the properties of the surface or presence of other ice species, along with the local gas density/pressure and gas temperature. For disks that are more edge on, it is sometimes possible to measure a sharp drop in submillimeter emission signifying the freeze-out boundary for a specific molecule.

While this does not provide a temperature ``map'' it still provides a key temperature data point at a specific vertical height. Depending on the disk physical structure and the excitation of the molecule of interest, however, care must be taken to ensure the drop in emission is not just an excitation effect in colder gas, and therefore a robust identification of the vertical snow line location would involve multiple spectral lines of the same species.

The second technique uses multiple transitions of optically thin lines to constrain the excitation temperature of the gas averaged over the observed column of gas. In disks, due to the high gas densities present, the excitation temperature is often equal to the gas kinetic temperature except in the upper tenuous layers or at very large radii, depending on a molecule's critical density. With high resolution observations of edge on disks such as Gomez's Hamburger \citep{Teague2020} or the Flying Saucer \citep{Dutrey2017}, one can directly measure the excitation temperature spatially and make temperature maps \citep[e.g.,][and references therein]{ruiz21}.A similar effort has been made for the massive inclined disks around Elias 2-27, as shown in \citep[][\emph{Paneque-Carre{\~n}o et al.,} to be subm.]{Paneque2021}, where the three main CO isotopologue lines and CN hyperfine structure lines have been analyzed, using a geometrical method, similat to that proposed by \cite{Pinte2018}. In less ideal inclination cases, such as TW Hya, one can still use this technique but the interpretation of the temperature's emitting region hinges on additional information, such as chemical models. For example \citet{loomis2018} was able to use the close spacing of the $J=12-11$ and $J=13-12$ $k$-ladder of CH$_3$CN to estimate a single band rotational temperature and column density for this species. Coupled with knowledge of the emitting layer of this molecule motivated from astrochemical models, they suggest that the temperature of the molecular layer ($z/r\sim0.3$) is as cool as 30 -- 40 K.

\begin{figure*}[ht]
      \centering
      \includegraphics[width=\hsize,trim={0cm 0.1cm 0cm 0cm},clip]{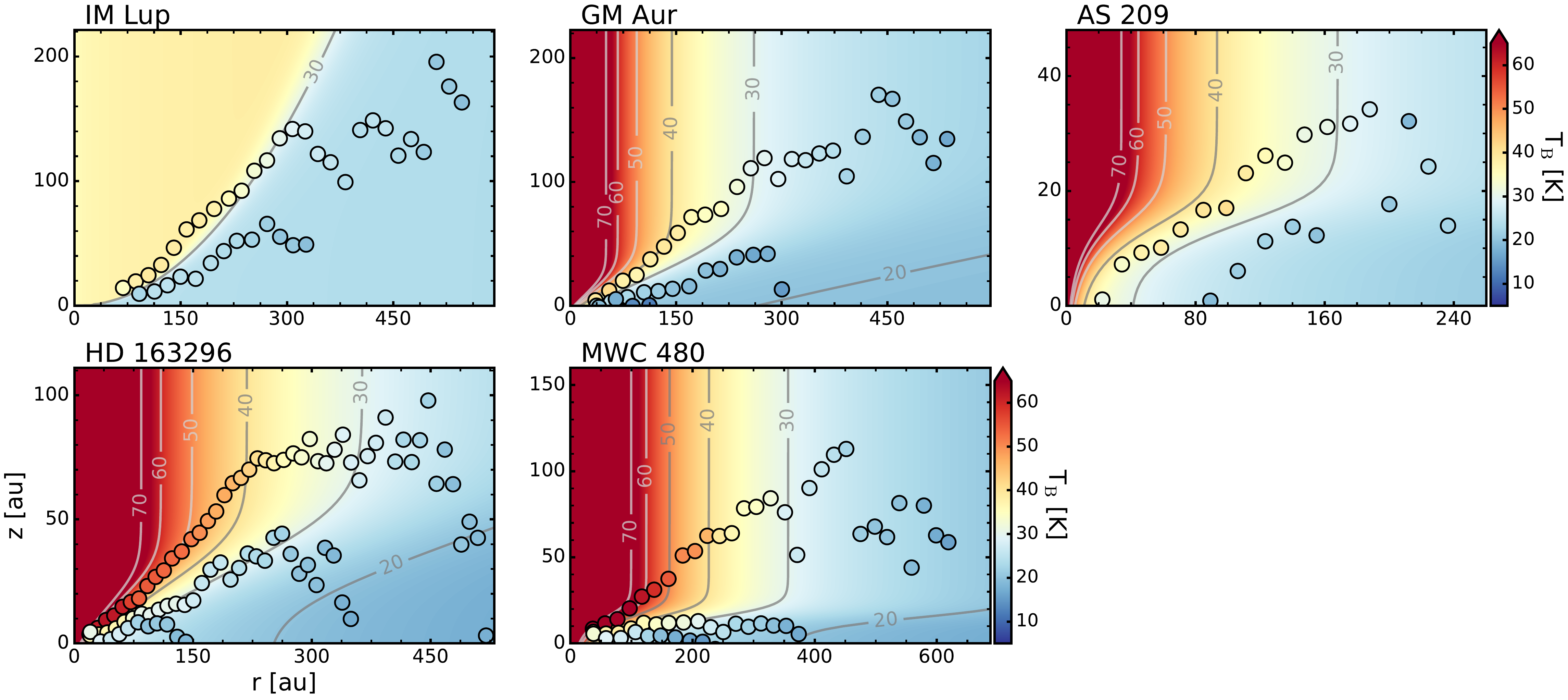}
      \captionsetup{}
      \caption{2D temperature distribution of CO, $^{13}$CO, and C$^{18}$O  (2-1) in all MAPS sources \citep{Law21_MAPS_surfaces_vertical_distributions}. The background color gradient shows the 2D temperature structure inferred from the data points (shown as circular points). Reproduced with permission of the authors.}
      \label{Fig:MAPS}
\end{figure*}

If the spectral line emission is optically thick, then the brightness temperature at the location of $\tau=1$ can be measured and interpreted as a gas temperature. For disks closer to edge on, the optical depth can be more murky, where the Keplerian velocity is seen more along the line-of-sight, correspondingly spreading out the line emission in frequency and reducing the overall optical depth. Nonetheless, very optically thick emission in disks, like for low J rotational lines of $^{12}$CO, can be used to provide strong constraints on the spatial temperature structure, as demonstrated by \citet[e.g.,][]{Schwarz2016,Pinte2018,Cleeves2018} and recently by \citet{Law21_MAPS_surfaces_vertical_distributions} as shown in Figure \ref{Fig:MAPS}.
For face-on disks, another option is to use multiple isotopologues whose emission is optically thick. This has been applied to TW Hya using the $J=3-2$ or $J=2-1$ rotational emission lines of $^{12}$CO, $^{12}$CO, and C$^{18}$O, where even  C$^{18}$O is optically thick in the inner tens of AU. Since these isotopologues have increasingly lower abundances relative to H$_2$, they become optically thick at different heights.\\ 

\textbf{The midplane} of the disk -- or the ``planet forming region'' is especially important to understand since it can drive both the composition of planetesimals (e.g., through radial snow line locations) and the details of the formation mechanisms for planets themselves. Techniques used to estimate the temperature of the midplane have overlap with those of the molecular layer, specifically localizing snow lines and using emission that becomes optically thick at heights corresponding to the midplane, like $^{13}$C$^{18}$O for the case of TW Hya \citep{Zhang2017}.

However, given that large dust grains sediment out and concentrate in this region, there are some unique tools that can be used. For example, millimeter dust grains which emit at centimeter to submillimeter wavelengths can have emission that becomes optically thick, especially in the inner disk. In this case, the brightness temperature of the emission reveals the dust thermal temperature, which is equal to the gas temperature in the dense midplane where frequent collisions equilibrate these two components temperatures \citep[e.g.,][]{woitke2015}. A challenge of this method that has become apparent over recent years is that disks are much more structured than previously thought. Thus while the continuum might be optically thick in narrow radial bands, there will be optically thin gaps between these bands that, when unresolved, result in under-estimates of the true dust temperature \citep[e.g.,][]{Andrews2018}. High spatial resolution observations and multi-wavelength observations are required to accurately pinpoint the continuum and confirm its opacity \citep[e.g.,][]{Kim2019}.

The continuum emission can reveal temperature information through other properties as well. For example,  \citet{lindaniel2020} find that temperature gradients can imprint on the observed polarization fraction because different polarization angles probe varying line of sight optical depths. Of interest, the proposed mechanism does not depend on the details of the alignment mechanism (e.g., magnetic or radiative), and instead simply requires grain alignment resulting in polarized flux. Conversions from polarized flux patterns to line of sight temperature gradients are provided in \citet{lindaniel2020}, describing the dependencies on factors such as disk inclination. Through radiative transfer modeling, this technique is applied to the young HH~212 disk, and warm ($\gtrsim45$~K) midplane temperatures are inferred \citep{lindaniel21}.

Another technique to estimate the midplane temperature is through using highly spatially resolved observations of an optically thick line like $^{12}$CO observed in an inclined disk. Through projection effects, the optically thick upper surface of the near side of the disk and the thick bottom surface of the far side provide constraints on the temperature differential between the layers. \citep{Dullemond2020}. This technique will primarily work for either warmer regions of a disk around a low mass star, where e.g., tracers like CO do not freeze out, or for disks around more massive stars that have very extended CO snow lines, as in the case of HD~163296.

\begin{summarybox}
      As of this chapter, there are not many cases where multiple techniques are applied to the same disk to make a cohesive structure verified by multiple techniques to estimate midplane temperature, molecular layer temperature, and so forth. It seems that for some low mass stars, the molecular layer may not be as strongly heated as some thermochemical models suggest, which may point to the variations in the presence or absence of cooling agents than what would be expected from standard interstellar cloud chemistry. Disks around Herbig Ae/Be stars indeed appear on average warmer in all layers \citep{Zhang2021}, but it is unclear how much of this can be attributed to the enhanced luminosities of these stars, or the fact that many of them have substantial structure, including large gaps that can change the penetration of radiation. Cross comparison studies using multiple techniques described here are essential to getting a better picture of disk vertical structure, which is in itself important for understanding planet formation mechanisms.
\end{summarybox}
\section{\textbf{SYNTHESIS AND OPEN QUESTIONS}}
\label{sec:synthesys}

\subsection{\textbf{Measurement of fundamental disk properties: open questions and implications}}
In Sec. \ref{sec:fund_prop} we have highlighted the main disk properties that set the disk structure and evolution from the initial formation of the disk all the way to planet formation. Despite the recent improvements in observational capabilities and modeling methods, it is clear that many open questions still remain and pave the way to future work.

For instance, on the topic of disk masses, large disk surveys suggest that Class II disks may not have enough solid material to build planetary cores, while younger disks may do. Would this conclusion still hold when all caveats on dust mass determination (see Sec. \ref{sec:dust_mass}) are considered? A similar question could be asked on the Class II gas content: is there enough gas mass in Class II disks to build gas giant planets, like Jupiter?

On the radial distribution of material within disks, could the surface density and temperature constraints be used to conclude anything on the stability of disks and the origin of dust substructures?
Also, which populations of dust grains - i.e. which grain sizes - would we expect to migrate, settle, and is this consistent with observations of e.g., thermal emission, polarization, dust traps?

One of the potential explanations for faint CO fluxes in disks is C and O elemental depletion. Do we have observational evidence that elemental abundances vary throughout the disk and do we understand the physical and chemical processes causing it? Are all our observational clues - e.g., CO-based disk outer radii measurements, turbulence levels, mid-IR spectra of HCN, C$_2$H$_2$, CO$_2$, water, sub-mm emission - consistent with the proposed scenarios?

Finally, which role does the midplane temperature structure, and accordingly the ice lines location,  play in setting mechanism(s) of planet formation and eventual planet composition? Does the imprint of an ice line survive the dynamical evolution that might accompany planet formation (e.g., pebble drift, planetesimal scattering), such that it can be utilized to identify a planet's original formation environment?
In this section, we try to address some of these open issues, and discuss some proposed ways forward.

\subsubsection{\textbf{A case study on $M_\mathrm{disk}$: TW Hya}}
\label{ex_TWHya}

\begin{figure*}[ht]
      \centering
      \includegraphics[width=14cm]{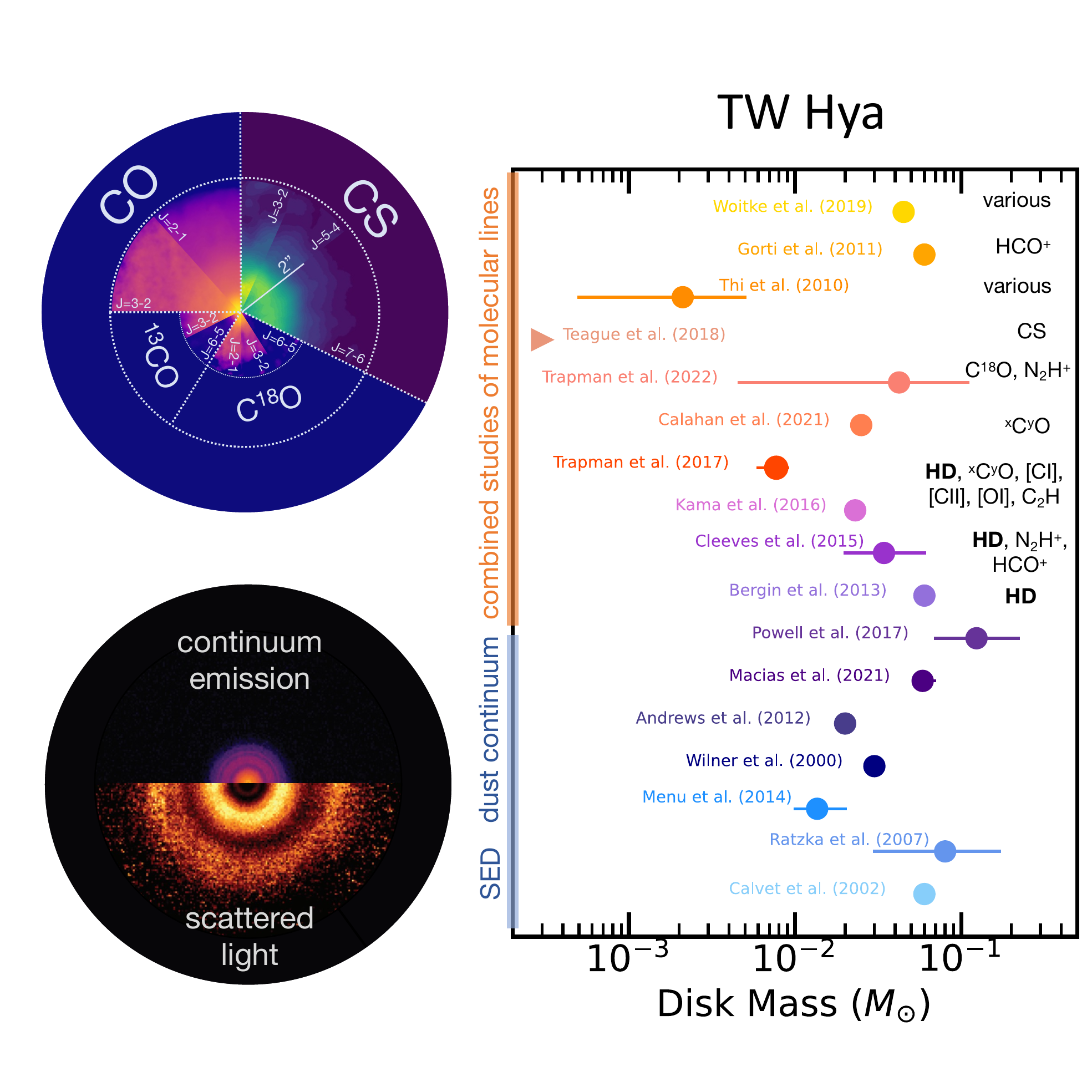}
      \captionsetup{}
      \caption{\emph{Left.} Collage of observations of TW Hya in different gas and dust tracers: CO isotopologues \citep{Calahan2021}, CS \citep{Teague2018}, scattered light from small dust \citep{vanBoekel2017} and sub-mm continuum from pebbles \citep{Andrews2016}. \emph{Right.} Measurements of the TW Hya disk mass from different tracers and studies (more details in Sec. \ref{ex_TWHya}. In some cases, where only the best fit result is reported in literature, we are unable to plot the actual uncertainty on the mass measurement (see circles with no error bars).
      }
      \label{Fig:TWHya}
\end{figure*}
TW Hya is arguably the most thoroughly studied disk, thanks to its proximity \citep[59.5 pc,][]{GaiaDR2} compared with other nearby SFRs ($> 100$ pc) and face-on inclination.  TW Hya has been observed at many different wavelengths, resolutions and sensitivities, and with different molecular and atomic tracers, which make it the best case study for the mass measurement ``problem.'' Moreover, this was the source with the first and most clear HD detection to date \citep{Bergin2013}.
Different measurements of the TW Hya disk mass are reported in the right plot of Fig.~\ref{Fig:TWHya}, while some of the tracers' images are shown on the left. This visual representation already shows how much the radial extent and the intensity of different tracers can vary, and consequently, that the final mass estimates will rely on large assumptions, extrapolations, and modeling efforts. It does not come as a surprise that the TW Hya disk mass measurements span two orders of magnitude (see right panel of Fig. \ref{Fig:TWHya}).

The most famous and possibly direct  TW Hya mass measurement\footnote{i.e., independent on chemistry, but still suffering from excitation and dust optical depth.} is based on the HD (1-0) line and is 0.06 M$_{\odot}$ \citep{Bergin2013}. Comparing the HD to the C$^{18}$O emission, an elemental gas phase carbon and oxygen deficiency of a factor of 10-100 has been inferred for this disk \citep{Hogerheijde2011,Favre2013,Du2015}. This deficiency is confirmed by physical-chemical modeling that aims to reproduce different atomic and molecular tracers, such as CO isotopologues, HD, [OI], [CI], [CII], and C$_2$H lines \citep{Kama2016}. Combining different tracers has allowed \cite{Kama2016} to better constrain the disk thermal structure, together with the elemental abundances and the disk mass, which turns then out slightly lower compared to the initial HD estimate: 0.023 M$_{\odot}$. This value is in line with previous studies combining CO, continuum and SED analysis \citep{Andrews2012,Menu2014,Cleeves2015,Woitke2019} as well as with the recent result by \cite{Calahan2021} that use resolved CO ALMA observations to constrain the temperature structure.
It is important to note that the modeling parameters can however vary between these different works. \cite{Andrews2012}, for instance, use a fixed CO mass fraction of $2\times 10^{-6}$ and a dust-to-gas mass ratio of 0.014; they use two grain populations (with dust sizes up to 1 $\mu$m and up to 1000 $\mu$m) and they have a fixed dust opacity at 870 $\mu$m of 3.4 cm$^2$g$^{-1}$. \cite{Menu2014} only use mid-IR to mm interferometry together with SED modelling. They obtain a dust disk mass of $1-2 \times 10^{-4} M_{\odot}$. They assume a grain size distribution from 0.01 to 10000 $\mu$m, and the opacity value not reported. \cite{Cleeves2015} use two vertically segregated dust populations, with sizes from 0.005 to 1/1000 $\mu$m with an opacity of 10 cm$^2$g$^{-1}$ at 870 $\mu$m. The gas-to-dust ratio is between 75 and 100, the dust mass is $4\times 10^{-4} M_{\odot}$ and the CO abundance is constrained to be $10^{-6}$.
Finally, \cite{Woitke2019} find a dust mass of $10^{-4} M_{\odot}$, with a gas-to-dust mass ratio of 450; the grain size distribution is from 0.0011 to 5700 $\mu$m with an opacity of 11.1 cm$^2$g$^{-1}$. No elemental depletion is assumed to set the CO abundance, that comes directly from the chemical network. Despite taking different assumptions, all these works seem to converge to a dust muss of $\sim 10^{-4} M_{\odot}$. However, more recently \cite{Macias2021} estimated a total dust mass between $7.5 \times 10^{-4} M_{\odot}$ and $10^{-3} M_{\odot}$, combining high angular resolution ALMA data at 0.87, 1.3, 2.1, and 3.1 mm, and considering the effects of optical depth, self-scattering, as well as a varying density profile, dust size distribution, and temperature profile.

The first TW Hya HD-based mass measurement published by \cite{Bergin2013} is generally used as the "real" gas mass and used as an anchor to constrain the gas-to-dust mass ratio or the elemental C and O abundances \citep[see e.g., ][]{Macias2021}. As discussed in Sec. \ref{sec:gas_mass} and in Sec. \ref{Mdisk_new}, translating the HD flux to the total disk gas mass is not straightforward, as line excitation and optical depth play an important role. It is therefore critical to constrain the uncertainties on the disk mass measurement, also when using HD lines. In this context, building up on the result by \cite{Bergin2013} and \cite{Kama2016}, \cite{Trapman2017} have implemented CO isotope-selective processes as well as a simple deuterium chemistry in their modeling and found that the disk mass can be between 0.006 and 0.009 M$_{\odot}$, depending on the disk vertical structure. The improved chemical modeling leads to an extrapolated mass that is lower compared to the one found by \cite{Kama2016}. Such detailed studies, combining different tracers, converge to a disk mass measurement within a factor of ten (0.006-0.06 M$_{\odot}$) and point to a less dramatic but still present deficiency of carbon and oxygen, i.e., volatile C and O elemental abundances need to be reduced, but only by a factor of few tens. Less recent works provide mass measurements that are on the low-hand side, see e.g. \cite{Thi2010} where the analysis, supported by SED and physical-chemical modeling, was based on the [OI] and CO line emission only, considering also $^{13}$CO and [CII] upper limits.

\subsubsection{\textbf{A way forward for constraining $M_\mathrm{disk}$}}
\label{Mdisk_new}
Combining different tracers seems the safest way for constraining disk masses, but which are the most effective combinations of tracers? HD is often considered the most robust gas mass tracer, but needs to be combined with a good characterization of the disk vertical and thermal structure, given the high temperatures needed for excitation of its ground state transition \citep{Trapman2017}. This can be provided by HD observations themselves, when two transitions are observed \citep[$J=1-0$ and $J=2-1$,][]{Kamp2021} and/or by resolved CO observations \citep[see e.g.,][]{Calahan2021}. In this context the implications of the cancellation of SPICA have been particularly severe for the disk community, as measuring disk masses observing HD ($J=1-0$ and $J=2-1$) lines with SAFARI in a large and unbiased sample of disks was one of the main goals. Such an HD survey will help to ``calibrate'' the CO-based masses, in order to accurately interpret the results from the ALMA disk surveys. A spectroscopic infrared space mission such as SPICA would be key.

In the absence of HD observations, another possible approach is to calibrate CO in a data driven way, by seeking trends between different but related molecules, whose emission is detectable by available facilities. If bulk gas is disappearing, one would expect other non-CO gas tracers to also decrease
in line flux. On the contrary, if the CO abundance is changing \citep[e.g., through conversion to species like organics,][]{Reboussin2015,Yu2017,Eistrup2018}, chemistry will be radically altered and the effect will be evident in the emission of other species. A combination of molecules that may help to disentangle the impact of changes in CO abundance ([CO]/[H$_2$]), changes in gas-phase C/O ratios, and changes in the overall gas mass is presented below. A tracer of the absence of CO and presence of H$_2$ gas is N$_2$H$^+$. This molecule is in fact readily destroyed in the presence of CO and forms from H$_2$ \citep[e.g.,][]{Bergin1997,Tafalla2004,vantHoff2017,Anderson2019}. Observations of CO faint disks in the 5 Myr old Upper Sco region indeed show high [N$_2$H$^+$]/[CO] ratios compared to younger, 1-2 Myr old disks \citep{Anderson2019}. These results imply that these older disks are more H$_2$ gas-rich than previously estimated from CO. On the same line, the recent work by \cite{Trapman2022} shows that a combination of N$_2$H$^+$ and C$^{18}$O lines can be used to reduce the uncertainty on the measurements of disk  masses, expecially if the cosmicray ionization rate can be better constrained.

C$_2$H and other hydrocarbons' abundances are also highly sensitive to gas phase C/O ratios \citep[e.g.,][]{Bergin2016,Cleeves2018,Miotello2019,Fedele2020,Bosman2021}. C/O is a useful proxy for understanding whether the disk gas is O or C dominated or of interstellar composition. Many observed disks appear to be missing oxygen in their molecular gas \citep{Hogerheijde2011,Du2017}, leading to C/O elevated well above the solar ratio, even $> 1$ \citep[][]{Miotello2019,Bosman2021}. Similarly, HCN is also a C/O sensitive molecule \citep{Najita2013,Cleeves2018} and provides an additional key nitrogen constraint since it is one of the most abundant forms of nitrogen that is still readily observable. Finally, [CI], [CII], and [OI] observations, when available, have been proven themselves as invaluable anchors to the volatile carbon, oxygen abundances and C/O ratios \citep[see e.g.][]{Kama2016}. The mid-IR is hence an interesting spectral window to measure the C/O ratio in the planet forming regions of disks \citep{Najita2013,Woitke2018}.

Deeper gas surveys of large and representative sample of disks, where this type of tracers are targeted, e.g. HD in the far-IR and/or ALMA multiple species (hydrocarbons, CO, N$_2$H$^+$), are still missing and would be needed in order to really probe the disks’ planet-forming potential by measuring reliable disk gas masses.

\subsubsection{\textbf{Implications of mass estimates for disk evolution and planet formation}}
\label{Disk_ev}

Together with the decline of gas accretion rate with time, the basic prediction of viscous accretion theory is the existence of a linear relation, with a coefficient
of order unity, between the mass accretion rate onto the central star and
the total disk mass \citep[see e.g., ][]{Hartmann1998}. Given the complexity of gas mass measurements, and the low number of disks for which this can be done homogeneously, dust masses multiplied by the canonical gas-to-dust ratio of 100 are generally used for studies of disk evolution in large samples of objects. Mass accretion rates are typically measured from the ultraviolet excess.
Combining the results of surveys conducted with VLT/X-Shooter and ALMA in the Lupus SFR, \cite{Manara2016} were able to test this basic prediction of viscous accretion theory for the first time. They found a correlation between the mass accretion rate and the disk dust mass, with a ratio that is roughly consistent with the expected viscous timescale when assuming a gas-to-dust ratio of 100. A similar result was also found in Chamaeleon I by \cite{Mulders2017}. Surprisingly however, no correlation is found when comparing mass accretion rates with CO-based gas masses. This is partly due to the lower detection rates of CO lines, but possibly also due to the fact that C and O depletion is at play at different levels in different disks within a single SFR (see Sec.~\ref{sec:gas_mass}). Deeper and unbiased ALMA surveys of CO emission in protoplanetary disks as well as a robust calibration of CO-derived disk masses are needed to further study this relation.

Disk dust masses obtained with ALMA disk surveys can also be compared with the masses of confirmed exoplanets, in order to explore when and how planets form in disks. \cite{Manara2018} find that exoplanetary systems masses are comparable or higher than the most massive Class II disks. A refined version of this study, accounting for observational selection and detection biases, has been carried out by \cite{Mulders2021}, who find a mitigated discrepancy between disk and exoplanets masses. Only if the planet formation efficiency was $\geq 100\%$, which is however much higher than predicted by planet-formation models especially if planets grow through pebble accretion \citep[see][and reference therein]{Tychoniec2020}, the discrepancy would be resolved. \cite{Manara2018} conclude that such result can be interpreted in two ways: either the cores of planets have formed very rapidly (in less than 1 Myr) and a large amount of gas is expelled on the same timescale from the disk, or disks are continuously replenished by fresh material from the environment.

After subtracting the envelope contribution, \cite{Tychoniec2020} have used ALMA (1.1 – 1.3 mm) and VLA (9 mm) continuum observations of embedded Class 0 and I disks in the Perseus SFR to provide a robust estimate of dust disk masses. The dust masses derived from the VLA data are higher than the typical Class II disk dust masses, reducing the tension with the observed exoplanetary system masses. If planet formation starts in the Class 0 stage with an efficiency of $\sim 15\%$, or in the Class I stage with an efficiency of $\sim 30\%$, the solid content in observed giant exoplanets can be easily obtained. One caveat is that dust opacity may evolve from class 0/I to Class II disks, while \cite{Tychoniec2020} assume the same opacity for all stages. If the dust opacity was higher at earlier times, for example if dust grains were much smaller than 1 mm in size, or the composition was different, the mass estimates of such young disks would be lower.

In summary, predictions on the planet formation efficiency from planet-formation models combined with observational comparisons between disk dust masses and exoplanet cores seem to show that Class II disks do not have enough solid mass to build planetary cores. This tension is reduced, if brighter Class 0/I disks are considered. There may be enough solid material at earlier stages, but a combination of higher temperature and higher dust opacity may be biasing this result. A still open question is whether there is still enough gas mass at the Class II stage to build gas-giants, like Jupiter. This unfortunately relies on our ability to measure disk gas masses and gas surface densities.

An extensive discussion on how measured disk bulk properties can be used for constraining disk evolution and planet formation can be found in the PPVII chapter by Manara et. al.

\subsection{\textbf{Dust properties: continuum vs constraints from polarization and implications for grain growth and migration as signposts of planetesimal formation}}

\label{Sec:scattering}

\label{sec:dust_properties}

The traditional way of constraining size of dust grains is to measure the spectrum at millimeter wavelengths, as explained in \autoref{sec:dust_mass}.
The opacity slope $\beta$, where $\kappa_\mathrm{abs} \propto \lambda^{-\beta}$ with $\kappa_\mathrm{abs}$ being the absorption opacity and $\lambda$ being the wavelength, is constrained to be as low as 0 to 1 in most of Class II disks \citep{Beckwith1991, Ricci2010a, Ricci2010b, Ansdell2018, Tazzari2020a}, as well as brown dwarf disks \citep{Ricci2014}.
This has been generally interpreted as existence of millimeter to centimeter large dust grains \citep{Miyake1993, Pollack1994}.

A trend of a radially increasing spectral index was already found in the pre-ALMA era \citep{Isella2010,Miotello2012,Perez2012,Tazzari2016,Menu2014,Perez2015}, which was interpreted as the grain size being larger in the inner part of the disk. The spatially resolved ALMA observations since then confirmed this trend \citep{CarrascoGonzalez2019,Macias2021,Tazzari2021} and show the trend that the spectral index is enhanced at gap regions \citep[][]{Tsukagoshi2016,Sierra2021} , which is the sign of small dust grains at gaps, consistent with the dust trapping scenario with a planet at the gap.

Recently, a completely different method of measuring the grain size has been proposed, which uses the scattering-induced polarization at millimeter wavelengths.
It has been long believed that the millimeter-wave polarization is dominated by intrinsic emission of elongated dust grains aligned with magnetic fields, which tells us the orientation of magnetic fields at midplane \citep{Cho2007, Bertrang2017}.
However, recent theoretical progress reveals that scattering-induced can also be a major mechanism of millimeter-wave polarization \citep{Kataoka2015}.
This is called as self-scattering of thermal dust emission.
Since the self-scattering polarization is detectable only if the grain size is $\sim$ (wavelength)$/2\pi$, multi-wavelength polarization allows us to constrain the dust grain size (see \autoref{fig:opacities}).

ALMA polarimetric observations show that several disks emit self-scattering polarization at 0.9 mm wavelengths \citep{Stephens2014, Stephens2017, Hull2018, Sadavoy2019, Dent2019, Bacciotti2018}.
Especially, HL Tau and DG Tau show the self-scattering at 0.9 mm and it decreases at 3 mm observations \citep{Stephens2017, Bacciotti2018, Harrison2019}, which is a strong evidence of that the grain size is up to $\sim 100$ micron \citep{Kataoka2016, Kataoka2017}.
The results are not consistent with the spectral index analysis, which infers the millimeter to centimeter sizes.

The scattering cannot be ignored even for the continuum emission.
Optically thick emissions are generally blackbody radiation.
However, if the dust albedo is high enough, the emission would be fainter than the blackbody radiation \citep[see][and \autoref{fig:opacities}]{Miyake1993, Birnstiel2018, Liu2019, Zhu2019, Sierra2020}.
This explains the spectral index lower than two.
In addition, this also allows us to constrain the grain size because the wavelength dependency of the dust albedo, or the scattering opacity, is a strong function of grain size.
The SED analyses have revealed that the grain size reaches 300 micron in the inner disk of TW Hya \citep{Ueda2020}, and 1-2 millimeter on HL Tau \citep{CarrascoGonzalez2019}.

To summarize the recent progress of measuring the grain size: while the spectral index analyses prefer the grain size millimeter to centimeter, the self-scattering polarization prefer the grain size of 100 micron, and the scattering-induced intensity reduction prefers 1 millimeter in grain size.
Now we will discuss how we can solve the discrepancy.

One idea is to keep the 100 micron dust grains but using optically thick components to explain the low spectral index.
\citet{Ricci2012} has proposed that spatially resolved optically thick components significantly reduce the spectral index, which misleads to the interpretation of large dust grains.
Disk size-luminosity analyses also supports the idea \citep{Tripathi2017}.
This provides one way to solve the inconsistency.
\citet{Lin2020} showed a model of dust continuum of HD 163296, where dust grains are made of 100 micron size and they are optically thick at the ring locations.
This naturally explains the low spectral index at rings as well as the brightness in polarization \citep{Dent2019}.

\citet{Ueda2021} have shown, that the discrepancy can be solved with  extremely settled large grains, requiring turbulence parameters of $\alpha \lesssim 10^{-5}$. \cite{Sierra2021} favor the large grain solution and propose longer wavelengths, higher resolution observations to confirm this.

Dust porosity and dust shape are other key factors to investigate.
Albedo of extremely fluffy aggregates (e.g., with a volume filling factor of $<< 0.01$) are shown to be too low to emit polarization \citep{Tazaki2019}.
In addition, the porous structure also reduced the intrinsic polarization of elongated thermal dust emission \citep{Kirchschlager2019}.
The detection of both self-scattering polarization and intrinsic polarization at least rules out extremely fluffy dust aggregates.
Furthermore, non-spherical grains have higher efficiency of polarization even with millimeter size dust grains \citep{Yang2019, Bertrang2020}.
In summary, further modeling -- especially with regards to wavelength-dependent behavior of non-spherical grains -- is needed to prove if they are the solutions to the dust size conundrum.

\subsection{\textbf{Disk vertical structure, turbulence and viscosity: implications for mixing and radial transport in view of planet composition}}
As discussed in Section~\ref{sec:vertstruct}, it is now well accepted that disks are vertically geometrically thick in the gas and small grains, and have relatively flatter vertical distributions in large, $>$mm-sized, grains, as expected by the dynamics of particles in a gaseous medium. As the midplane is the site of planet formation, it is important to understand the vertical concentration of gas and dust populations and their physical and chemical implications. For example, minerals that have experienced high temperature processing have been found in parent bodies at large distances from the Sun \citep[e.g.,][]{boss2008,Kleine20}, leading to the questions like: how did they get there? Were they processed at the warm/hot disk surface and mixed down? Or were they radially mixed? And what about volatiles? How much is redistributed when dust grains coated with icy-mantles settle to the midplane \citep{Krijt2018}? And what implications does transport have for the measured isotopic ratios present in primitive solar system bodies, along with planet forming disks \citep[e.g.,][]{lyons2005,lichtenberg21}?

These questions around disk mixing (and many more) are intractably tied to our understanding of gas and dust bulk transport. Physical processes that transport material radially, enabling, for example, accretion onto the star also generate hydrodynamic turbulence. Since PPVI, our understanding of the nature of disk viscosity and its implications for turbulence have been put back into the forefront. Given the challenge of detecting accretion through the disk directly (mainly relying on the measured accretion rates onto the central star), evidence for transport mechanisms through the disk is often inferred through the search for non-Keplerian motions that are the underlying source or producer of viscosity itself. Sensitive observations with very high spectral resolution have allowed the field to search for such non-Keplerian motions, including turbulent broadening, in spectral lines. These methods have been employed both in the inner ($<5$ au) disk using IR observations of broadening in the CO bandhead \citep{carr2004}, as well as the outer disk at submillimeter wavelengths \citep{Hughes2011}, including with higher spatial resolution observations with ALMA \citep{Flaherty2015,Teague2016,Flaherty2018,Flaherty2020}.

While significant broadening has been measured in the inner disk \citep[at $2\times$ the local sound speed][]{carr2004}, in most cases, the degree of non-Keplerian broadening in the outer ($>20$~au) disk measured is low or at the limits of the observations \citep[see][and references therein]{Flaherty2020}. The magnitude of turbulent broadening is moreover at odds with viscous disk accretion theory if the source of viscosity is magnetohydrodynamic turbulence \citep{Simon2015,Simon2018} as is classically assumed \citep{balbus1991}. These findings have resulted in changes to the paradigm of disk accretion, but also have important implications for mixing and our broader interpretation of the disk environment.

      {\em Mixing and dust growth:} Collisions between small dust grains are the starting point of the assembly of planets. Turbulence provides the necessary initial ``kick'' to enable the aggregation of particles; however, the details of the process are highly complex. The outcomes of collisions are highly dependent on a variety of material and environmental properties \citep[e.g.,][]{windmark2012,windmark2012b,blum2018}. Facilitating growth requires some non-zero level of random motions that is neither too weak nor too strong such that particles do not fragment on impact \citep[][and references therein]{dominik07,Testi2014}. These motions, required for growth, impact the vertical distribution of dust. Observations of light scattering by small dust grains in the infrared and optical, including a recent survey of \citet[][see also Figure~\ref{Fig:IMLup}]{Avenhaus2018}, have indicated that grains are present at substantial distances from the midplane, $\sim2$ gas scale heights in some cases. The challenge of these observations' interpretation, however, is that they provide little information about the quantity of dust at these elevations since very little small dust is required before the disk becomes optically thick. Nonetheless, the {\em presence} of dust at high altitude seems to suggest that there must be a force effectively keeping dust from settling out or continuously lofting dust upward. High spatial resolution observations of millimeter emission, tracing the large dust grains, as a function of azimuthal angle suggest that the midplane is geometrically thin, with thicknesses as small as $\sim1$ au at 100 au \citep{Pinte2016}. This would suggest that the midplane dust population is not being strongly stirred, at least in a handful of disks \citep[see also][]{Grafe2013,Villenave2020}. These results, alongside the finding of little to no non-Keplerian line broadening in the molecular gas, put interesting constraints on the magnitude of disk turbulence. So then how do the small grains probed by scattered light remain aloft? The settling time for small grains at 50 au from 2 scale heights is 600,000 years, and 1 scale height is 2.7 million years. Therefore {\em stopping} small grains is itself not the primary issue. Instead, collision times are expected to be rapid. If collisions are driven by even weak turbulence, equivalent to $\alpha = 10^{-4}$, grains will collide in a relatively short time, taking only 1000s to 10,000s of years at 50~au, for a variety of grain sizes and porosities. To prevent growth and subsequent settling, the timescale for collisions must be increased, perhaps by removing all but a very small amount of dust, $<< 1\%$, however this would need to be made consistent with the observed scattering surfaces. Alternatively, if a property of the dust itself can provide coagulation, such as grain charging \citep{Okuzumi09,Okuzumi2011,matthews2012}, that would enable a sustained presence of small grains. Another explanation is that our measurements of turbulence primarily probe more quiescent layers such as the warm molecular layer \citep{Flaherty2020} and the midplane \citep{Pinte2016}, and perhaps the rarefied surface is actually turbulent, enabling frequent destructive collisions that would keep dust grains small. Sensitive, high spectral resolution observations of surface tracing species are necessary to test this hypothesis.

      {\em Mixing and volatile composition:} The degree of vertical mixing has important implications for the composition of the planet forming midplane and our interpretation of measured solar system abundances, especially isotopes. Chemical models have been developed to simultaneously incorporate turbulent mixing and chemistry \citep[e.g.,][]{heinzeller11,semenov11, Furuya2013,yang2013}. So far, the challenge of validating models -- with or without mixing -- mainly lies in  model complexity itself and in degeneracies with input parameters. Unique signatures of chemical variations due to mixing and/or transport have yet to be concretely identified and observationally verified. One interesting puzzle that has emerged in recent years is the observation of bright emission from small hydrocarbons  \citep{Bergin2016,Cleeves2018,bergner19,Miotello2019,bosman21}. Chemically, formation of these species is favored under super-solar C/O conditions, especially if C/O $\gtrsim$ 1 \citep{Cleeves2018,Bosman2021}. Observations of sulfur bearing molecules find a similar requirement of high C/O \citep{legal2021}. The vertical distribution of the high C/O layer, as traced by the hydrocarbons, is moreover elevated at $>2$ scale heights based on direct imaging, chemical modeling, and excitation arguments \citep[][]{kastner15,Cleeves2018,cleeves2021,guzman21}. While the outer disk is sufficiently cold, even at the disk surface,  for H$_2$O to freeze out with CO present in the gas, models suggest that H$_2$O freeze-out alone is not sufficient if the medium is UV irradiated since photodesorbed water can reintroduce oxygen in the gas phase and suppress the C$_2$H abundance. Instead, oxygen must be ``removed'' from the surface, including H$_2$O as well as CO, to leave sufficient excess C to form hydrocarbons. Mechanisms to remove oxygen could involve a ``cold finger'' type of mechanism where freeze out of vapor diffusing or mixing across the vertical snow surface leaves the gas ``dry'' of volatile oxygen \citep[e.g.,][]{Kama2016}. Another mechanism is through the differential dynamical evolution of the solids, where growth, settling, and radial drift redistribute not only the grain itself but also any ice coatings on the grain \citep[see, e.g.,][]{Krijt2018,oberg16,piso2015}. Large scale surveys of disk chemistry combined with settling and dust evolution tracers along with accretion tracers would be beneficial to better understand the chemical implications of mixing for forming planets.

      {\em Mixing and isotopic composition:} Observations of solar system bodies find peculiar isotopic composition in both refractory solids and volatile ices. Of particular interest are H, O, C, and N isotopes, whose isotopologues can be observed in interstellar and circumstellar environments, facilitating comparison with our solar system \citep{ceccarelli14}. While we will leave the details of the measurements and chemical pathways to the PPVII chapter ``The Isotopic Link from the Planet Forming Region to the Solar System,'', here we briefly discuss how mixing can impact our interpretation of isotopic anomalies.

The primary fractionation pathways are mass-dependent or photo-selective. Mass-dependent fractionation becomes important at low temperatures where the energetic differences in a reaction having a heavier versus a lighter isotope become important. Photo-selective fractionation occurs when molecules are more readily destroyed in the presence of dissociating radiation, primarily due to self-shielding. As a result, isotopologue ratios have been widely used as a probe of the original formation environment -- temperature or radiation field -- of a molecular species.

For example, high fractions of deuterated water and organic material in our solar system point to formation in a cold ($<30$~K) environment \citep{robert2000}. Variations in the deuterium fraction, especially within the same parent body, have been used as evidence to suggest that transport occurred between regions of varying temperature, however caution must be taken as there are many different fractionation pathways that have different temperature barriers \citep{roueff13} resulting in different deuteration levels for molecules originating in the same environment \citep{Thi2010i,Furuya2013,cleeves2016b}. Isotopologues comprised of the more massive O, C, and N tend to be less influenced by mass-dependent fractionation, and instead are fractionated more efficiently by photon-driven processes. Specifically, CO and N$_2$ are dissociated by particular UV wavelengths \citep{vanDishoeck88,heays14}. Thus when a UV irradiated slab of CO or N$_2$ effectively runs out of photons, these species are no longer destroyed and are ``shielded'' from further destruction. Less abundant isotopologues such as $^{13}$C$^{16}$O or $^{15}$N$^{14}$N are less effective at shielding, and therefore these molecules are destroyed, enhancing the gas with the less abundant isotope. These atoms are then incorporated into the formation of other molecules creating enhancements in rarer isotopologue bearing species.

Both mass dependent and photon driven fractionation ratios have been measured in our solar system and observed directly in protoplanetary disks \citep{huang17,guzman17,oberg21}. As found in these works, deuterium ratios are often higher than both what is found in our solar system and in pre-stellar cores. Moreover, using excitation arguments, \citet{oberg21} found that the emission was arising from the warm surface and not the midplane, inconsistent with active chemical fractionation in the midplane. This observation leads to the question -- are these high degrees of fractionated molecules mixed into the midplane and incorporated into larger bodies? And if they are not, how are the high magnitudes of volatile fractionation achieved? Are they inherited from an earlier stage, such as in the cold cloud? The photon-driven fractionation processes also add another interesting piece to the puzzle. Similar to deuterium, high amounts of $^{18}$O and $^{15}$N bearing molecules are observed in disks, exceeding solar system measurements as well. In this case, the surface fractionation is expected given the high UV fields present there, while little active fractionation is expected in the UV shielded midplane. Is this also a result of mixing, now bringing material from the surface down \citep{lyons2005}? It remains unclear how much mixing changes the midplane vs. surface isotopologue ratios, and observations of fractionated molecular carriers closer to the midplane are required to understand how much surface to midplane transport plays a role in either enhancing or decreasing midplane fractionation.

\section{\textbf{SUMMARY}}
\label{Sec:summary}

In the past few years, with ALMA becoming fully operational, our field has seen a revolution: not only spectacular morphological features of protoplanetary disks have been discovered \citep[see e.g.,][]{ALMAPartnership2015,Andrews2018,oberg21}, but also large surveys of disks have become affordable. This has allowed us to make the first steps towards a deep knowledge of the bulk disk's fundamental physical characteristics as a population. Disk surveys of continuum and CO isotopologues emission, in some of the closest star-forming regions, have allowed us for the first time to independently analyze the presence of gas and solids in a statistically significant number of disks \citep[][]{Ansdell2016,Ansdell2017,Ansdell2018,Pascucci2016,Barenfeld2016,Eisner2016,Manara2016,Long2017,Miotello2017,Mulders2017,Tychoniec2018,Cazzoletti2019,Cieza2019}.\\

However, these observations have highlighted a problem that is common to all regions: Class II disks seem to be not sufficiently massive to generate the observed exoplanetary population \citep[e.g.,][]{Ansdell2016,Manara2018,Mulders2021}. This ``missing mass proble'' seems to affect both the solid and gaseous components of disks, but not usually in the same way.
Are disks intrinsically mass poor, and planet formation has already fully occurred, or is a large fraction of the bulk mass obscured to our detection methods?

More specifically, \emph{dust masses} as derived from sub-mm thermal emission of dust grains, may be affected by the unknown dust opacity, temperature and optical depth of the continuum emission. Furthermore, we do not know how much mass is hidden in the planetesimals which may or may not have formed and that are not traceable by thermal emission.
Scattering, even at (sub-)millimeter wavelengths, has now proven to play a major role in our interpretation of thermal continuum and polarization observations \citep{Liu2019, Zhu2019,Kataoka2015}. Regions where the thermal emission appears optically thin, could in fact be highly optically thick due to dust scattering. Dust polarization observations are dominated by dust scattering and find a maximum grain size of $\sim 100\, \mu$m, in tension with traditional size measurements from the dust spectral index \citep{Beckwith1991,Kitamura2002,Testi2014} and with theoretical predictions of pressure trapping of dust grains and planetesimal formation.

In the past few years, \emph{gas masses} have been mostly derived from CO isotopologues, whose line emission was too faint to be detected by the - unfortunately - too shallow ALMA disk surveys, for the bulk of the disk population. This observation can be interpreted as a sign of fast gas dispersal or of quick chemical evolution. The second hypothesis is supported by the observationally derived mass accretion rates, which would not be sustainable in the first scenario \citep{Manara2016}. Furthermore, \emph{Herschel}-PACS observations of the HD fundamental line in a few bright disks showed that CO-based gas masses can be up to two orders of magnitude smaller than HD-based disk masses \citep[e.g.,][]{Bergin2013,Favre2013}. For other sources the ratio between HD and CO derived masses is a factor of a few \citep{McClure2016,Trapman2017,Kama2020}. This potential inconsistency can be explained by locking up of volatiles as ices in larger bodies, depleting the gas in oxygen, and thus leading to low observed CO fluxes. This hypothesis is supported by observations of other molecular species such as C$_2$H and N$_2$H$^+$ \citep[see e.g.,][and references therein]{Cleeves2018,Miotello2019,Anderson2019}. Unfortunately, with the cancellation of the next far-IR space observatory SPICA, we will not have any observatory within the next 2 decades capable to detect HD lines in a large unbiased sample of disks and calibrate CO-based disk masses \citep{Kamp2021}. {Alternative more focused mission/instrument concepts have been proposed to fill the niche of detecting HD in representative samples of young disks. While the HIRMES project on SOFIA \citep{Richards2018} also got cancelled, the NASA/MIDEX-class mission concept OASIS is still under discussion as of the writing of this chapter \citep{Walker2021}.} Dynamical constraints on the disk gas mass have been proposed and are promising, as they do not rely on the emission of a specific tracer \citep{Rosenfeld2013,Veronesi2021,Hall2020,Terry2021,Powell2017,Powell2019}. Nevertheless, this type of analysis is only applicable to the most massive and brightest disks and are affected by other caveats, such as for example the need of an accurate measurement of the disk age, and the complications due to the presence of strong substructures. All these methods, as well as other described throughout this chapter, have their flaws and rely on different types of assumptions.

The \emph{surface density distribution} is another property that is fundamental for determining planet formation and the architecture of planetary systems. Depending on which disk evolution process is dominant, i.e.\ viscous evolution, disk winds, or external processes,
the distribution of material as a function of the distance from the star will be considerably different \citep{Morbidelli2016}. Reliable observational measurements of $\Sigma$ as function of radius are therefore key. Continuum observations at moderate spatial resolution are used to constrain the slope of $\Sigma_\mathrm{dust}$ using different model approaches \citep[e.g.][]{Tazzari2017,Tsukagoshi2016}. Generally, $\Sigma$ appears flatter in the inner disk and steepens at larger radii. How these measurements compare with the $\Sigma_\mathrm{gas}$ profile, which is what governs viscous evolution, is not yet clear. \cite{Miotello2017} have shown that CO isotopologues spatially resolved observations can be used to constrain the slope of $\Sigma_\mathrm{gas}$. The inner ($<5$ au) disk $\Sigma$ has been constrained from CO ro-vibrational lines for disks with cavities using VLT/CRIRES \citep[e.g.,][]{Carmona2014,Carmona2017}. Surface density measurements have however the same uncertainties as mass measurements when it comes to the normalization, i.e., the dust opacity and the abundance of CO with respect of H$_2$. A determination of $\Sigma (R)$ for a large number of disks is still not available yet, but would be key.

Most established theories describe disk evolution as driven either by viscosity, or by magnetically supported winds, which lead to different predictions on the evolution of the \emph{outer radius}, with the former predicting that $R_\mathrm{out}$ should expand with time. Measuring the disk gas radius is therefore a valuable test for disk evolution processes.
Optically thick $^{12}$CO lines are used to trace the gas outer radius as they are bright also in the outer disk, where the surface density drops. Disks in different star forming regions have been targeted with $^{12}$CO observations, but CO radii ($R_\mathrm{CO}$) have been extracted only for a fraction of them \citep[][]{Ansdell2018,Boyden2020}, and this has limited so far any statistical study of disk radii. A large fraction of disks have not been resolved nor detected in CO by ALMA surveys. A possible interpretation is that disks with faint CO fluxes may in fact be radially compact \citep{Pietu2014,Hendler2017,Woitke2011,Boneberg2018,vanTerwisga2019,Miotello2021}. Deeper CO observations of fainter disks are missing and are urgently needed. On the other hand, continuum observations require shorter integration times and can be used to determine the radial extent of mm-sized particles, to constrain $R_\mathrm{mm}$ \citep{Tazzari2017,Tripathi2017}. For a small sample, where both CO and mm continuum data exists, $R_\mathrm{CO}$ appears to be universally larger than $R_\mathrm{mm}$ by a factpr 2-3, as it comes naturally from CO line optical depth \citep[see e.g.,][]{Facchini2017}. There are however a few disks that have mm radii that are a factor 4-5 smaller than those of CO, pointing to efficient radial migration \citep{Facchini2019}. In order to test viscous evolution. Recently, some effort was put into modeling CO emission, dust evolution and growth, and comparing it with available observations \citep{Trapman2020,Trapman2021,Toci2021}. The results, however, do not yet fully agree with each other. Larger samples are required, together with modelling, to understand the size distribution of disks, its evolution with time and the processes that drive it.

The radial \emph{temperature profile} $T(R)$ is key for understanding the disk structure and evolution as well as the formation mechanism(s) of planets \citep[see, e.g., review of][]{armitage07}. It is hard to directly extract from observations: gas lines probe the temperatures only in the layers where they emit, generally above the midplane \citep[e.g.,][]{Fedele2013,vanderWiel2014,Schwarz2016,Pinte2018}. Ice-line tracers are sometimes used as midplane temperature calibration points, such as N$_2$H$^+$, however interpretation likely requires detailed thermo-chemical modeling \citep{vantHoff2017,Schwarz2019}. The continuum requires multi-wavelength data to break degeneracies between dust properties (grain opacity, size distribution) and temperature \citep{Kim2019}. Often radial temperature gradients are only indirectly inferred from radiative transfer modeling \citep[e.g.,][]{Cieza2017}. New approaches are emerging thanks to the combined high spatial and spectral resolution of ALMA such as extracting brightness temperatures between the CO emission lobes \citep{Dullemond2020}. Much can be improved upon of our knowledge of the radial temperature profile, especially at intermediate and outer disk regions. Unlike dust observations, the lack of high quality - i.e., deep and with high angular resolution - gas data and the need for coherent multi molecule/line data sets. Recent results from the ALMA Large Program MAPS, despite being restricted to only 5 of the brightest disks, go in the other direction and pave the way to similar studies in a larger sample of objects.

Observations of the disk \emph{vertical structure} show clear indication of dust settling. While small $\mu$m-sized grains trace the flared surface of the gas, the mm-dust resides in the midplane \citep[][]{Lagage2006,Avenhaus2018,Pinte2016}. Similarly to scattered light emission, $^{12}$CO lines originate in the flaring disk surface, while CO isotopologue lines can trace deeper to the colder layers near the CO-ice surface \citep[][]{Schwarz2016,Dutrey2017,Pinte2018,Teague2020,Paneque2021,Law21_MAPS_surfaces_vertical_distributions}. The vertical thermal and chemical structure is predicted by thermo-chemical disk models, which provide information beyond CO. However, the spatially resolved molecular line observations to compare these detailed models to is only slowly building up with ALMA. It is also becoming evident that dust optical depth cannot be neglected when interpreting gas emission, even in the outer disk \citep[][]{Cleeves2016,Rab2020}; the vertical structure of gas and dust is thus intertwined and hard to interpret independently. Finally, recent measurements have found turbulence in the outer disk (beyond 50~au) to be generally subsonic \citep{Hughes2011, Guilloteau2012,Flaherty2015,Flaherty2018,Teague2016} \\

In conclusion, with ALMA becoming operational and the arrival of a new generation of  high-contrast optical and IR instruments, the field of planetary system formation has undergone transformative change. As reported in this chapter, our knowledge has dramatically improved in many aspects of the disk structure and evolution, especially because different theories are nowadays testable with observed quantities. On the other hand, many paradigms have been questioned by recent discoveries and have opened many questions, that are not yet closed and have been highlighted, especially in Sec. 3. Finally, most of the observational effort has been put to date on detailed studies of bright and massive disks (see e.g., Sec.~\ref{ex_TWHya}), while the bulk of the disk population is heavily understudied. We lack deep CO surveys of large samples of disks in a representative number of star forming regions. Molecular surveys of a limited number of disks are just starting and, on the other hand, ALMA archival data is often too heterogeneous (spatial resolution, frequency coverage) to address some of these major open issues. We are confident that in the coming few years observational and modeling efforts will be put into studying the fundamental properties of the bulk disk population allowing a proper understanding of disk evolution and planet formation.
\\

\noindent\textbf{Acknowledgments} The authors wish to thank Feng Long and Olja Panic for sharing the Cha I measurements and the Herbig disk data respectively used in Fig. \ref{Fig:g2d}. This research has made use of NASA's Astrophysics Data System and of \texttt{adstex} (\url{https://github.com/yymao/adstex}).

\bibliography{bibliography}

\end{document}